\documentclass[a4paper,11pt]{article}
\usepackage{jheppub} % for details on the use of the package, please see the JINST-author-manual
\usepackage{lineno}
\usepackage{graphicx,subfigure}
\usepackage{float}
\usepackage{hyperref} % For hyperlinks
\usepackage[capitalize]{cleveref} % must be loaded after hyperref

\title{A systematic study of hydrodynamic attractors in periodically driven systems at weak and strong couplings}
\author[a]{Aleksas Mazeliauskas,}
\author[a]{Toshali Mitra,}
\author[b]{Sören Schlichting,}
\author[b]{Simon Schneider,}
\author[b]{Louis Nwakile Onwuka,}
\author[b]{and Martin Vrdoljak}

\usepackage[dvipsnames]{xcolor}
\affiliation[a]{Institut für Theoretische Physik, Universität Heidelberg, D-69120 Heidelberg, Germany}
\affiliation[b]{Fakultät für Physik, Universität Bielefeld, D-33615 Bielefeld, Germany}

\emailAdd{a.mazeliauskas@thphys.uni-heidelberg.de}
\emailAdd{mitra@thphys.uni-heidelberg.de}
\emailAdd{sschlichting@physik.uni-bielefeld.de}
\emailAdd{simon.schneider@uni-bielefeld.de}
\emailAdd{louis.onwuka@uni-bielefeld.de}
\emailAdd{martin.vrdoljak@uni-bielefeld.de}
 
\date{\today}

\abstract{
We study equilibration and hydrodynamic response in conformal systems undergoing periodic expansion and contraction along one spatial direction. Using strongly coupled holography (AdS/CFT), weakly coupled kinetic theories (relaxation-time approximation, scalar $\phi^4$ theory and the Yang--Mills sector of QCD effective kinetic theory) and hydrodynamics (Navier--Stokes, Müller--Israel--Stewart (MIS) and rBRSSS), we investigate the shear response to a periodic drive of varying amplitude $A$ and frequency $\omega$, both analytically in linear response and numerically. We find that, at the linear level, $A\ll 1$, periodically driven systems approach a cyclic attractor that in general differs from Navier--Stokes behaviour. This attractor is universal among kinetic theories and accurately described by MIS theory: for slow driving, $\omega \tau_\pi \ll 1$, where $\tau_\pi$ is the shear relaxation time, it approaches the Navier--Stokes limit, whereas for fast driving the shear stress saturates at its non-interacting (free-streaming) value.
In strongly coupled holographic systems, the periodic drive excites non-hydrodynamic quasinormal modes, which produce a characteristic ringing in the approach to the cyclic attractor, in contrast to the purely exponential approach found in kinetic theories. This difference is a direct consequence of the distinct non-hydrodynamic spectra of the two classes of theories. Moreover, unlike in kinetic theory, the holographic response is sensitive to the curvature terms of second-order hydrodynamics.
Over longer times, non-linear viscous heating, which is quadratic in the drive amplitude, raises the temperature and thereby lowers $\omega\tau_\pi$, since $\tau_\pi\sim 1/T$. This induces a slow drift of the cyclic attractor, which eventually converges towards Navier--Stokes hydrodynamics at asymptotically late times.}
\begin{document}

\maketitle

\section{Introduction}

The unexpectedly early applicability of relativistic hydrodynamics in heavy-ion collisions, long before the quark-gluon plasma reaches local thermal equilibrium, is a cornerstone of our understanding of these collisions \cite{Busza:2018rrf,Schlichting:2019abc,Berges:2020fwq}. This success is underpinned by the concept of hydrodynamic attractors~\cite{Heller:2015dha}: after a short transient, the evolution of the energy-momentum tensor loses sensitivity to the details of the initial state and follows a universal curve, which only at late times is described by a low-order truncation of the hydrodynamic gradient expansion \cite{Florkowski:2017olj,Soloviev:2021lhs,Jankowski:2023fdz,Enss:2026jps}. The study of such attractors has since grown into an active area of research, connecting a variety of microscopic theories, from kinetic theory to holography, with mathematical tools such as the theory of transseries and resurgence \cite{Florkowski:2017olj,Aniceto:2018bis,Jankowski:2023fdz}.

More recently, these ideas have been extended beyond the monotonic Bjorken expansion that is characteristic of heavy-ion collisions. For instance, attractors have been studied in Hubble-like expanding systems related to cosmology \cite{Du:2021fok}, in Gubser flow \cite{Denicol:2018pak,Behtash:2019qtk,Dash:2020zqx,Mitra:2024zfy}, in fluctuating hydrodynamics \cite{Chen:2022ryi}, and in more general phase-space formulations \cite{Heller:2020anv,Spalinski:2025ngd}. In particular, they have received growing attention in ultracold atomic gases with a time-dependent scattering length, which is equivalent to an isotropic expansion \cite{Fujii:2018aik}. In such systems, attractors have been proposed for monotonic drives \cite{Fujii:2024yce}, for early-time expansion-driven loss of initial-state information \cite{Heller:2025yxm}, and for periodic expansion and contraction \cite{Mazeliauskas:2025jyi}. Such oscillating expansion histories may also be relevant to the post-merger phase of neutron-star mergers \cite{Gavassino:2023xkt} and to cold-atom experiments \cite{Xie_2026}. Periodic driving provides a qualitatively different regime for probing the limits of hydrodynamic behaviour and the robustness of the attractor, since the system is never simply left to relax towards equilibrium.

In this paper, we present a comprehensive study of hydrodynamic attractors in a conformal system undergoing periodic expansion and contraction along one spatial direction, which excites the shear response of the energy-momentum tensor. Although this setup does not correspond to the longitudinal expansion of a heavy-ion collision, it allows us to isolate the conditions under which attractor behaviour emerges, to contrast the different theoretical limits, and to study the distinction between the loss of sensitivity to the initial state and the onset of validity of a low-order truncation of the hydrodynamic gradient expansion, as emphasised in a recent review \cite{Enss:2026jps}.

To span the range from weak to strong coupling, we employ several complementary frameworks. On the macroscopic side we use first-order Navier--Stokes hydrodynamics as well as second-order Müller--Israel--Stewart (MIS) theory \cite{Muller:1967zza, Israel:1979wp} and rBRSSS hydrodynamics~\cite{Baier:2007ix,Romatschke:2017ejr,Keegan:2015avk}, which provide the standard benchmarks for non-equilibrium fluid dynamics. At the microscopic level we consider three kinetic descriptions: the relaxation-time approximation (RTA), chosen for its analytic transparency \cite{Denicol:2014mca,Strickland:2017kux}, a scalar $\phi^4$ theory \cite{Jeon:1994if,Jeon:1995zm}, and the Yang--Mills sector of QCD effective kinetic theory (EKT) \cite{Arnold:2002zm,Kurkela:2015qoa,Kurkela:2018oqw,Du:2020dvp}, which captures the elastic and inelastic processes of weakly coupled gauge theories. Finally, to reach the limit of infinite coupling we use AdS/CFT holography \cite{Maldacena:1997re,Gubser:1998bc,Witten:1998qj}, which provides a dual gravitational description of the non-equilibrium plasma~\cite{Chesler:2008hg,Heller:2011ju}.
This multi-model approach allows us to follow the attractor across the amplitude and frequency regimes of the periodic drive, to identify the universal features that persist across otherwise disparate theories, and to understand how the approach to the attractor is shaped by the underlying microscopic dynamics.

The paper is organised as follows. In \cref{sec:setup} we describe the periodic expansion setup and introduce the macroscopic and microscopic theories used in this work. In \cref{sec:analytic} we derive the linear response of each theory in terms of its retarded shear-stress response function, whose value at the driving frequency fixes the cyclic attractor and whose analytic structure determines how the attractor is approached.
In \cref{sec:numerics} we compare these expectations with numerical simulations, first in the small-amplitude limit for all theories, then for the non-linear response in RTA kinetic theory, and finally for the long-time evolution, where viscous heating slowly lowers $\omega\tau_\pi$. In contrast to Bjorken flow, where the Knudsen number decreases in time, the driving frequency here is an external control parameter, so the approach to a cyclic attractor can be followed both inside and outside the hydrodynamic regime. We summarise our findings in \cref{sec:conclusions}, and details of the holographic calculation are collected in \cref{app:holoappendix}.

\section{Setup}\label{sec:setup}
We consider a system undergoing a one-dimensional expansion, as described by the following metric
\begin{align}\label{eq:metric}
    ds^2 = -dt^2 +  dx_\perp^2 + a(t)^2 d \eta^2
\end{align}
where  $a(t)$ is the time-dependent scale factor.
Unlike the Bjorken case $a(t)=t$, which is flat Minkowski space in Milne coordinates, the metric \cref{eq:metric} is curved for a general $a(t)$. Its only non-vanishing curvature components are $R^{t\eta}{}_{t\eta}=\ddot a/a$, such that $R_{tt}=-\ddot a/a$, $R_{\eta\eta}=a\ddot a$ and $R=2\ddot a/a$. Due to the explicit time dependence of the scale factor, one has the non-trivial Christoffel symbols: $\Gamma^{t}_{\eta \eta}= a \dot a $, and $\Gamma^{\eta}_{\eta t}=\Gamma^{\eta }_{t \eta}= \frac{\dot a}{a}$. Moreover, $\sqrt{-g}=\sqrt{-\det g_{\mu\nu}} = a$. Since the system is homogeneous in the coordinates of \cref{eq:metric}, it remains at rest, as described by the local rest frame $u^\mu=(1,0,0,0)$. However, the system is subject to a one-dimensional expansion, with an expansion rate $\theta = \nabla_\mu u^\mu= \dot a/a$, where $\nabla_\mu$ is the covariant derivative.

\begin{figure}
    \centering
    \includegraphics[width=0.5\linewidth]{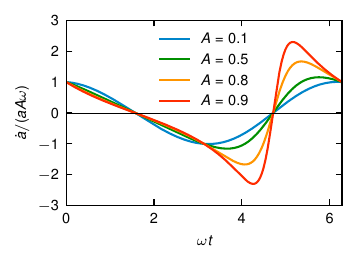}
    \caption{Expansion rate $\theta(t)$ normalised to $A\omega$ for different values of the amplitude $A$.}
    \label{fig:expansion_rate}
\end{figure}

While the case $a(t)=t$ corresponds to Bjorken flow, we will consider a periodic drive that starts with $a(t=0)=1$ and $\theta(t=0)=A\omega $, as given by 
\begin{align}
    a(t)&=1+ A \sin(\omega t)\Theta(t)\;,\label{eq:pert}\\
    \theta(t)&=\frac{A\omega \cos(\omega t)}{1+ A \sin(\omega t)}\Theta(t)\;,
\end{align}
such that $0\leq A<1$ controls the amplitude, while $\omega$ corresponds to the frequency at which space-time contracts and expands. We show in \cref{fig:expansion_rate}  how the expansion rate $\theta$ varies over the course of one period, for different amplitudes.
While for $A \ll 1$ the behaviour is close to harmonic, one sees that for large amplitudes $A$, the system rapidly transitions from contracting to expanding at $\omega t= 3\pi/2$, where the behaviour of the expansion rate becomes divergent in the limit $A \to 1$.

In this work, we consider different macroscopic and microscopic descriptions of the medium driven by expansion in \cref{eq:metric}. We will focus on the time evolution of the energy-momentum tensor $T^{\mu\nu}$, which for
conformal systems can be decomposed as
\begin{equation}
T^{\mu\nu} = e u^\mu u^\nu + P_\text{eq}\Delta^{\mu\nu} + \pi^{\mu\nu}
\end{equation}
where $e$ is the energy density, as determined by $T^{\mu\nu}u_\nu = - e u^\mu$, $P_\text{eq}= e/3$ denotes the thermodynamic pressure, $\Delta^{\mu\nu}= g^{\mu\nu}+ u^\mu u^\nu$ is the projector orthogonal to $u^\mu$, and $\pi^{\mu\nu}$ denotes the symmetric traceless shear-stress tensor, orthogonal to the fluid velocity, $\pi^{\mu\nu}u_\nu=0$. Due to the symmetries of the system, the energy-momentum tensor $T^{\mu}_{~\nu}= \text{diag}(-e, P_T, P_T, P_L)$  is diagonal in the coordinates of \cref{eq:metric} and there is no bulk pressure for conformal theories. One thus finds  \begin{align}
    T^{\mu}_{~\nu}=\text{diag}(-e, P_\text{eq}-\pi/2, P_\text{eq}-\pi/2, P_\text{eq}+\pi)\;,
\end{align}
where for notational simplicity we defined
\begin{equation}
    \pi \equiv \pi^\eta_\eta.
\end{equation}

Based on the conservation equation $\nabla_{\mu}T^{\mu\nu}=0$ for the one-dimensional expansion given by the metric in \cref{eq:metric}, the energy density evolves according to
\begin{align} \label{eq:eom}
    &\partial_t e (t)=- \frac{\dot a}{a} \left(e+P_\text{eq} + \pi\right)
\end{align}
A macroscopic hydrodynamic description of the system is therefore achieved by providing the constitutive equations for $\pi^{\mu\nu}$ in terms of hydrodynamic fields $u^\mu$ and $e$ as well as their derivatives \cite{Baier:2007ix,Romatschke:2009kr,Romatschke:2017ejr}. In microscopic theories, such as kinetic theory, the shear stress $\pi^{\mu\nu}$ can be obtained microscopically from moments of the single-particle distribution function $f$, which obeys a Boltzmann equation \cite{Denicol:2021clh}. In holographic theories, there is no quasi-particle interpretation of the medium, and instead the evolution of $T^{\mu\nu}$ is obtained by solving the dual Einstein equations in $d+1$ dimensions and extracting the boundary stress tensor from the near-boundary geometry \cite{Chesler:2013lia,Balasubramanian:1999re,deHaro:2000vlm}.

Below we discuss the different microscopic and macroscopic evolution models in more detail. In all cases, we will consider initial conditions where the system is in thermal equilibrium at initial time $t=0$, such that there is no initial shear stress, $\pi(t=0)=0$.

\subsection{Navier--Stokes theory (NS)}
Hydrodynamics provides a universal long-time and long-wavelength description of interacting systems close to thermal equilibrium. Hydrodynamic constitutive relations for the dissipative fluxes, such as $\pi^{\mu\nu}$, can be derived based on an expansion in gradients of the thermodynamic variables. Neglecting dissipative fluxes corresponds to ideal hydrodynamics, which is the zeroth order of the expansion. The first order in this expansion is commonly referred to as Navier--Stokes hydrodynamics, for which the constitutive relation for the shear-stress tensor takes the following form \cite{Romatschke:2017ejr,Denicol:2021clh} 
\begin{equation}
    \pi^{\mu\nu}_\text{NS}=  -\eta \left( \Delta^{\mu\alpha} \Delta^{\nu\beta} \left( \nabla_\alpha u_\beta + \nabla_\beta u_\alpha \right)  - \frac{2}{3} \Delta^{\mu\nu} \nabla_\lambda u^\lambda \right)
\end{equation}
where $\eta$ is the shear viscosity. Specifically, for the expansion of the form \cref{eq:metric} the shear stress in the Navier--Stokes limit is given by
\begin{align}
\pi_\text{NS}=-\frac{4}{3} \eta \frac{\dot{a}}{a}\;,\label{eq:NSpi}
\end{align}
and one can expect to recover this result for any system which is slowly evolving in time and remains close to equilibrium, i.e.\ in the limit of small amplitude $A \ll 1$ and small frequency $\omega \tau_\pi \ll 1$.

\subsection{Second-order hydrodynamics (rBRSSS and MIS)}
It was realised early on that the relativistic Navier--Stokes equations are not causal and can be numerically unstable~\cite{Muller:1967zza,Israel:1979wp}. Even though there are by now stable and causal versions of first-order relativistic hydrodynamics~\cite{Bemfica:2020zjp,Kovtun:2019hdm}, the most common practice is still to promote the shear-stress tensor to a dynamical variable which relaxes to its Navier--Stokes value over a relaxation time $\tau_\pi$. Since this formally amounts to an inclusion of terms which are of second order in the derivatives of hydrodynamic fields, it is then only consistent to include all second-order terms, and different schemes to carry out a systematic expansion have been established in the literature~\cite{Denicol:2012cn}.

Specifically for an irrotational conformal fluid\footnote{For a general conformal fluid, the BRSSS equation contains two additional second-order terms, $-\frac{\lambda_2}{\eta}\pi^{\langle\mu}_{\lambda}\Omega^{\nu\rangle\lambda}$ and $\lambda_3\,\Omega^{\langle\mu}_{\lambda}\Omega^{\nu\rangle\lambda}$, involving the vorticity tensor $\Omega^{\mu\nu}$~\cite{Baier:2007ix}. These terms vanish for the irrotational flow considered here, $u^\mu=(1,0,0,0)$.}, the general form of the evolution equations to second order in derivatives has been determined by BRSSS~\cite{Baier:2007ix}; it is usually referred to as rBRSSS and takes the form 
\begin{align} \label{Eq:BRSSS}
    \pi^{\mu\nu} = \pi^{\mu\nu}_\text{NS} -\tau_{\pi} \left( {}^\langle D \pi^{\mu\nu \rangle} + \frac{d}{d-1} \pi^{\mu\nu} (\nabla\cdot u)\right) &+ \kappa \left( R^{\langle \mu \nu \rangle} - (d-2) u_\alpha R^{\alpha \langle \mu \nu \rangle \beta} u_\beta \right) \nonumber \\ 
    &+ \frac{\lambda_1}{\eta^2} \pi^{\langle \mu}_\lambda \pi^{\nu \rangle \lambda} 
\end{align}
where $d=4$ is the number of space-time dimensions, $\kappa$ and $\lambda_1$ are second-order transport coefficients that describe the system's linear response to space-time curvature and non-linear response to shear stress, respectively. Since conformal theories have no intrinsic scale, the temperature dependence of all transport coefficients is fixed by dimensional analysis, and each coefficient is characterised by a single dimensionless constant, $\eta = C_\eta\,(e+P_\text{eq})/T$ (where $e+P_\text{eq}=Ts+\mu n$, such that $C_\eta=\eta/s$ at vanishing chemical potential), $\tau_\pi = C_{\tau_\pi}/T$, $\kappa = C_\kappa\,\eta/T$ and $\lambda_1 = C_{\lambda_1}\eta/T$. The values of these constants for the different theories studied in this paper are summarised in \cref{tab:coef}. Evaluating \cref{Eq:BRSSS} for the expansion of the form \cref{eq:metric} and considering only the $\pi = \pi^\eta_\eta$ component leads to
\begin{align} \label{eq:rbrsss_nice}
     \pi = \pi_\text{NS} - \tau_\pi \left(\partial_t \pi + \frac{4}{3} \pi \frac{\dot{a}}{a}\right) - \kappa \frac{2}{3}\frac{\ddot{a}}{a} +\frac{\lambda_1}{2\eta^2}\pi^2 .
\end{align}
By keeping only terms that are linear in deviations from equilibrium, i.e.\ dropping the non-linear terms $\propto \pi \frac{\dot{a}}{a}$ and $\propto \pi^2$, 
the above evolution equation for $\pi$ reduces to the familiar Müller--Israel--Stewart theory. Specifically for our case we have
\begin{equation}
\tau_\pi \partial_t\pi+\pi = \pi_{\text{NS}} - \kappa \frac{2}{3} \frac{\ddot{a}}{a} \label{eq:MISeom}
\end{equation}
where the shear stress $\pi$ relaxes to its Navier--Stokes value $\pi_{\text{NS}}$ on a timescale $\tau_\pi$, up to second-order gradient corrections associated with the response to the space-time curvature.

\begin{table}[h!]
\centering
\begin{tabular}{|c|c|c|c|c|}
\hline
Theory & $C_\eta$ & $C_{\tau_\pi}/C_\eta$ & $C_{\kappa}/C_{\eta}$ & $C_{\lambda_1}/C_{\eta}$ \\ \hline
RTA & 
$T\tau_R/5$ 
&  $5$  & 0 & 25/7\\ \hline
Yang--Mills EKT ($\lambda=10$)  & 0.56 & $5.2$ & 0 & 4.1\\ \hline
Scalar & $6930/\lambda^2$ & $6.10517$
& 0 &$ 6.13264$ \\ \hline
Holography & $1/(4\pi)$ & $ (4-2\log(2))$ & 4 & 2\\ \hline
\end{tabular}
\caption[Transport coefficients]{Values of the relevant first- and second-order transport coefficients, $C_\eta = \eta T/(e+P_\text{eq})$ and $C_{\tau_\pi} = T \tau_{\pi}$, $C_{\kappa}= T \kappa/\eta$, $C_{\lambda_1} = T\lambda_1/\eta$ for the different theories studied in this paper. The RTA values are taken from~\cite{Ambrus:2022vif}, where the relaxation time $\tau_R$ is a free parameter. The Yang--Mills EKT value of $\eta/s$ is taken from~\cite{Arnold:2002zm,Arnold:2003zc,Boguslavski:2024kbd}, the remaining Yang--Mills EKT values from~\cite{York:2008rr}, the scalar $\phi^4$ values from~\cite{Jeon:1994if,Jeon:1995zm,York:2008rr}, and the holographic values for $\mathcal{N}=4$ supersymmetric Yang--Mills theory from~\cite{Baier:2007ix}. In the kinetic theories $C_\kappa=0$~\cite{York:2008rr}; the non-zero, but parametrically suppressed, field-theory values of $\kappa$ are computed in~\cite{Moore:2012tc}.
 \label{tab:coef}}
\end{table}

\subsection{Kinetic theory}\label{sec:kinetic}
Effective kinetic descriptions provide a more microscopic description of the dynamics of weakly interacting many-body systems, where the evolution of the single-particle distribution function $f(x,\vec p)$ obeys the relativistic Boltzmann equation~\cite{cercignani2002relativistic,Du:2021fok}
\begin{equation}
\left[p^\mu \partial_\mu  - \Gamma^i_{\alpha \beta}p^\alpha p^\beta \frac{\partial }{\partial p^i}\right]f(x, \vec p) = - \mathcal{C}[f],
\end{equation}
where the index $i$ signifies that only spatial indices $(x,y,\eta)$ are involved, while $p^{0}$  is evaluated on-shell, i.e.\ $p^0 = \sqrt{(p^x)^2+(p^y)^2+a(t)^2 (p^\eta)^2}$ for massless quasi-particles. Specifically, for a homogeneous system in the coordinates of \cref{eq:metric}, we can simplify the Boltzmann equation to the form below, where for convenience we absorbed the factor $1/p^0$ into the definition of the collision kernel, $C[f]\equiv\mathcal{C}[f]/p^0$, which we will use in the following
\begin{equation}
\left[\partial_t - 2 \frac{\dot a}{a}p^\eta \frac{\partial }{\partial p^\eta}\right]f(t,p^{x},p^{y},p^\eta) = - C[f], 
\end{equation}
where partial time derivatives are to be understood at constant $p^{\eta}$. However, for all practical purposes it proves more insightful to consider $p_{\|}=a(t)p^{\eta}$ as the longitudinal momentum variable, which carries the physical dimension of a momentum such that e.g.\ $p_{\eta}p^{\eta}=a(t)^2 (p^{\eta})^2=p_{\|}^{2}$, and we will compactly denote $\vec{p}=(p_{x},p_{y},p_{\|})$  in the following. In terms of these variables the Boltzmann equation reads
\begin{equation}
\left[\partial_t - \frac{\dot a}{a}p_{\|} \frac{\partial }{\partial p_{\|}}\right]f(t, \vec{p})= - C[f],
\end{equation}
and, based on its solution, the relevant components of the energy-momentum tensor can then be evaluated as moments of the single-particle distribution
\begin{align}
    e(t)&=-T^{t}_{~t}(t)=\nu \int_{p}~p^{0}~f(t,\vec{p}) \\
    P_{\text{eq}}(t)+\pi(t)&=T^{\eta}_{~\eta}(t)=\nu \int_{p}~\frac{p_{\|}^{2}}{p^{0}}~f(t,\vec{p})
\end{align}
where we introduce the short-hand notation $\int_{p}=\int \frac{dp_x}{(2\pi)}\int \frac{dp_y}{(2\pi)}\int \frac{dp_{\|}}{(2\pi)}=\int \frac{d^3p}{(2\pi)^3}$ and $\nu$ denotes the number of degrees of freedom. When investigating the behaviour in kinetic theory, we consider three different types of collision kernels $C[f]$ specified in more detail below.

\subsubsection{Free streaming}
Before addressing different kinds of interactions, we briefly note that in the absence of interactions, i.e.\ $C[f]=0$, the Boltzmann equation can be solved analytically using the method of characteristics, yielding the solution
\begin{align*}
    f(t,p_x,p_y,p_{\|})= f(t_0,p_x,p_y, p_{\|} a(t)/a(t_0) ),
\end{align*}
where physically, the appearance of the factor $p_{\|} a(t)/a(t_0)$ signifies the red/blue shift of the longitudinal momentum in the expanding space-time \labelcref{eq:metric}.

\subsubsection{Relaxation-time approximation (RTA)}

The simplest kinetic description is the relaxation-time approximation (RTA), in which the full collision integral is replaced by a relaxation term. Physically, the RTA describes a system in which the distribution function relaxes towards the local equilibrium distribution $f_\text{eq}$ uniformly in momentum, i.e.\ all momentum modes approach equilibrium exponentially on a single relaxation time scale $\tau_R$, without resolving the details of the underlying microscopic scattering processes. Despite its simplicity, the RTA captures the essential interplay between free streaming and equilibration, and serves as an analytically tractable benchmark for more realistic kinetic theories. The collision kernel takes the form
\begin{equation}
\mathcal{C}[f] =\frac{-u_\mu p^\mu  \left(f(t, \vec p)-f_\text{eq}(t,\vec p )\right)}{\tau_R(t)},
\end{equation}
where the local rest-frame velocity $u^{\mu}(x)$ and equilibrium temperature $T(x)$ that determine the equilibrium distribution $f_{\text{eq}}(t,\vec{p})$ are determined from the Landau matching condition $T^{\mu}_{~\nu}(x) u^{\nu}(x)=-e(T(x)) u^{\mu}(x)$ to satisfy energy-momentum conservation. Specifically, for a homogeneous system at rest, we have $u^{\mu}(x)=(1,\vec{0
})$, such that $C[f]=\mathcal{C}[f]/p^0=\left(f(t, \vec p)-f_\text{eq}(t,\vec p )\right)/\tau_R(t)$, and only need to determine the time-dependent equilibrium temperature $T(t)$ from the equation of state $e(T)=\nu \frac{\pi^2}{30} T^4$. We consider a conformal theory for which the relaxation time is determined as $\tau_R(t) T(t)=5 \eta/s = \text{const}$, with a constant ratio of shear viscosity to entropy density. Due to the simplicity of the collision kernel in the relaxation-time approximation, an implicit formal solution can be obtained (cf.~\cite{Du:2021fok}). In practice, however, we employ a moment expansion as in~\cite{Kamata:2020mka} to obtain numerical solutions for kinetic theory in the relaxation-time approximation.

\subsubsection{Yang--Mills kinetic theory}
Secondly, we consider the Yang--Mills sector of QCD, where the effective kinetic theory at leading order consists of elastic and inelastic processes~\cite{Arnold:2002zm}
\begin{equation}
    C[f](\vec{p}_1) = C^{2\leftrightarrow2}[f](\vec{p}_1) + C^{1\leftrightarrow2}[f](\vec{p}_1)
\end{equation}
In more detail, the collision integral for elastic $gg\leftrightarrow gg$ scattering is given by
\begin{equation}
    C^{2\leftrightarrow2}[f](\vec{p}_1) = \frac{1}{2\nu_g}\frac{1}{2E_{p_1}} \int d\Pi_{2 \leftrightarrow 2} \left| \mathcal{M}^{gg}_{gg}(\vec{p}_1, \vec{p}_2 | \vec{p}_3, \vec{p}_4) \right|^2 F^{gg}_{gg}(\vec{p}_1, \vec{p}_2 | \vec{p}_3, \vec{p}_4)
\end{equation}
where $d\Pi_{2\leftrightarrow 2}$ denotes the phase-space integration measure
\begin{equation}
   d\Pi_{2\leftrightarrow 2} = \frac{d^3p_2}{(2\pi)^3 2E_{p_2}} \frac{1}{(2\pi)^3} \frac{d^3p_3}{2E_{p_3}} \frac{1}{(2\pi)^3} \frac{d^3p_4}{2E_{p_4}} (2\pi)^4 \delta^{(4)}(p_1+p_2-p_3-p_4)
\end{equation}
the statistical factor is
\begin{align}
   F^{gg}_{gg}(\vec{p}_1, \vec{p}_2|\vec{p}_3, \vec{p}_4) = &f(\vec{p}_1)f(\vec{p}_2)(1+ f(\vec{p}_3))(1+ f(\vec{p}_4)) \nonumber\\
   - &f(\vec{p}_3)f(\vec{p}_4)(1+f(\vec{p}_1))(1+ f(\vec{p}_2))
\end{align}
and the vacuum matrix element is given by
\begin{equation}
    \left| \mathcal{M}^{gg}_{gg}(\vec{p}_1, \vec{p}_2 | \vec{p}_3, \vec{p}_4) \right|^2 = 4g^4 d_A C_A^2\left(9 + \frac{(s-u)^2}{t^2} + \frac{(s-t)^2}{u^2} + \frac{(t-u)^2}{s^2}\right)
\end{equation}
which inside the medium is regulated using HTL screening for $t$- and $u$-channel processes, as discussed in~\cite{Boguslavski:2024kbd,Boguslavski:2025ylx}. Similarly, at leading order the collision integral for inelastic processes can be expressed as
\begin{align}
C^{1\leftrightarrow2}[f](\vec{p})
=\frac{1}{2}\int_0^1dz
\left[\frac{d\Gamma_{gg}^{g}}{dz}\big(|\vec{p}|,z\big)F_{gg}^{g}(\vec{p}|z\vec{p},\bar{z}\vec{p})-\frac{2}{z^3}\frac{d\Gamma_{gg}^{g}}{dz}\big(\frac{|\vec{p}|}{z},z\big)F_{gg}^{g}(\frac{\vec{p}}{z}|\vec{p},\frac{\bar{z}}{z}\vec{p})\right]
\end{align}
where $z$ and $\bar{z}=1-z$ are the collinear splitting/joining fractions, and the statistical factor is given by
\begin{align}
F_{gg}^{g}(\vec{p}_1|\vec{p}_2,\vec{p}_3)
=f(\vec{p}_1)(1+ f(\vec{p}_2))(1+f(\vec{p}_3))-f(\vec{p}_2)f(\vec{p}_3)(1+ f(\vec{p}_1))
\end{align}
The effective inelastic rate $\frac{d\Gamma_{gg}^{g}}{dz}$ is obtained within the framework of AMY~\cite{Arnold:2002zm}, and, as discussed in~\cite{Du:2020dvp}, corresponds to the overall probability of a single radiative emission/absorption over the course of multiple successive elastic interactions, reduced to an effective collinear rate by integrating over the parametrically small transverse momentum accumulated during the emission/absorption process.

Numerical solutions to the above kinetic equations are obtained by evolving discrete moments of the integro-differential equation as described in~\cite{Du:2020dvp}. 

\subsubsection{Scalar kinetic theory}
We also consider a scalar $\phi^4$ theory~\cite{Jeon:1994if,Jeon:1995zm}, which at leading order describes elastic scattering with a simple collision matrix element $\left|\mathcal{M}(\vec{p}_1, \vec{p}_2 |\vec{p}_3, \vec{p}_4) \right|^2=\lambda^2$, such that the collision integral takes the form
\begin{align}
 C^{2\leftrightarrow2}[f](\vec{p}_1) = \frac{1}{2}\frac{1}{2E_{p_1}} \int d\Pi_{2 \leftrightarrow 2} ~\lambda^2~F^{gg}_{gg}(\vec{p}_1, \vec{p}_2|\vec{p}_3, \vec{p}_4).
\end{align}
We note that in contrast to QCD/Yang--Mills kinetic theory, the leading-order scalar kinetic theory is particle-number conserving, such that a general equilibrium state is characterised by a temperature $T$ and a chemical potential $\mu$.

\subsection{Holography}\label{sec:Holography}
To study the strongly coupled regime, we employ holography: in the limit of large $N_c$ and large 't Hooft coupling, the AdS/CFT correspondence maps $\mathcal{N} = 4$ supersymmetric Yang--Mills theory onto classical Einstein gravity in an asymptotically anti-de Sitter space-time \cite{Maldacena:1997re,Gubser:1998bc,Witten:1998qj}. 
The near-boundary behaviour of the bulk
geometry encodes {the expectation values of single-trace operators in
the dual state}, which can be extracted from the on-shell gravitational action via
holographic renormalisation
\cite{deHaro:2000vlm,Bianchi:2001kw,Skenderis:2002wp,Balasubramanian:1999re}.

For the time-dependent boundary metric \cref{eq:metric}, we consider the dual five-dimensional bulk geometry
in ingoing Eddington--Finkelstein coordinates,
\begin{align} \label{Eq:HoloMetric}
    ds_G^2 = -\left(2/r^2 \right) dt\, dr - \mathcal{A}(r,t)\, dt^2
    + \mathcal{S}(r,t)^2 \left( e^{\mathcal{P}(r,t)} (dx^2 + dy^2)
    + e^{-2 \mathcal{P}(r,t)} d\eta^2 \right),
\end{align}
whose conformal boundary is located at $r = 0$, and the boundary metric is recovered following the holographic dictionary, $\lim_{r \rightarrow 0} (r^2/l^2)\, G_{\mu\nu} = g_{\mu\nu}^{(b)}$, where $\mu,\nu$
are the boundary indices. The bulk geometry satisfies the Einstein equations
\begin{align} \label{Eq:Einstein}
    R_{MN} - \frac{1}{2} R\, G_{MN} + \Lambda G_{MN} = 0,
\end{align}
where $\Lambda = -6/l^2$ is the cosmological constant and $l$ is the AdS radius, which we set to $l=1$
henceforth. The ansatz \labelcref{Eq:HoloMetric} is invariant under the residual
gauge transformation $1/r \to 1/r + \xi(t)$, which we fix by setting $\xi = 0$.
Solving the Einstein equations near the boundary yields the metric functions $\mathcal{A}$, $\mathcal{P}$, and $\mathcal{S}$ as a series expansion in the radial coordinate $r$,
\begin{align} \label{Eq:NearBdyExp}
    \mathcal{A}(r,t) &= \frac{1}{r^2} - \frac{2a''}{3a} + r^2 A_4
    + \frac{r^2 \log r}{9a^2}\left(\frac{a'^2 a''}{a} + \frac{a''^2}{2} - a' a'''\right)
    + \mathcal{O}(r^3), \\
    \mathcal{P}(r,t) &= -\frac{2}{3}\log a - \frac{2r a'}{3a}
    + \frac{r^2\left(2a'^2 - a a''\right)}{6a^2}
    + \frac{r^3\left(a a' a'' + 3a^2 a''' - 12a'^3\right)}{54a^3} \nonumber \\
    &\quad + r^4 P_4 + \frac{r^4 \log r}{24a}\left(a^{(4)} - \frac{5a''^2}{3a}
    - \frac{5a' a'''}{3a} + \frac{5a'^2 a''}{3a^2}\right) + \mathcal{O}(r^5), \label{Eq:NearBdyExp1} \\
    \mathcal{S}(r,t) &= \frac{a^{1/3}}{r} + \frac{a'}{3a^{2/3}} - \frac{r\, a'^2}{9a^{5/3}}
    + \frac{r^2\left(5a'^3 - 3a a' a''\right)}{81a^{8/3}} + \mathcal{O}(r^3), \label{Eq:NearBdyExp2}
\end{align}
where $a \equiv a(t)$ and primes denote time derivatives. The
logarithmic terms reflect the conformal anomaly in $d=4$ which vanishes when the boundary
metric is conformally flat. The near-boundary analysis leaves the normalisable modes
$A_4(t)$ and $P_4(t)$ undetermined. However, $P_4(t)$ is related to $A_4'(t)$ by the Einstein constraint, which
is the boundary energy-momentum conservation,
\begin{align} \label{Eq:Constraint}
  A_4' &= -\frac{4a'}{3a}\left(2P_4 + A_4\right)
  + \frac{a' a''}{36a^3}\left(\frac{5a'^2}{a} - \frac{61a''}{9}\right) \nonumber \\
  &\quad - \frac{{a'''}}{18a^2}\left(a'' {+}
  \frac{109a'^2}{18a}\right)
  + \frac{a'}{108a^5}\left(48a'^4 + 17a^3 a^{(4)}\right){.}
\end{align}

In holographic renormalisation \cite{Balasubramanian:1999re,deHaro:2000vlm},
power-law divergences of the on-shell action are canceled by local counterterms, and
the logarithmic divergence associated with the anomaly by a counterterm proportional
to $\log r$. Here we use the minimal subtraction scheme, i.e.\ we add no finite counterterms \cite{Balasubramanian:1999re,deHaro:2000vlm}. 
This gives
\begin{align} \label{Eq:bdystresstensor}
    e &= -\frac{N_c^2}{4\pi^2}\left(\frac{3A_4}{2} + \frac{a''^2}{9a^2}
    - \frac{5{a'''} a'}{36a^2} + \frac{5a'^2a''}{36a^3}\right), \\
    P_T &= \frac{N_c^2}{4\pi^2}\left(-\frac{A_4}{2} + 2P_4 - \frac{7a^{(4)}}{144a}
    + \frac{19a''^2}{432a^2} - \frac{a'^4}{3a^4}
    + \frac{79{a'''} a'}{432a^2} - \frac{5a'^2a''}{144a^3}\right), \\
    P_L &= \frac{N_c^2}{4\pi^2}\left(-\frac{A_4}{2} - 4P_4 + \frac{7a^{(4)}}{72a}
    - \frac{25a''^2}{216a^2} + \frac{2a'^4}{3a^4}
    - \frac{49{a'''} a'}{216a^2} - \frac{5a'^2a''}{72a^3}\right)
   {.}
\end{align}
$A_4$ determines the energy density and $P_4$ the pressure anisotropy, up to local terms fixed by $a(t)$. The values of $A_4(t)$ and $P_4(t)$ are fixed by the full bulk
evolution. We obtain them numerically by
evolving the Einstein equations in the characteristic formulation
\cite{Chesler:2013lia} from an initial profile $\mathcal{P}(r,t_0)$ and $A_4(t_0)$,
extracting $P_4(t)$ from the near-boundary behaviour of $\mathcal{P}$
and integrating \labelcref{Eq:Constraint} {to obtain $A_4(t)$}.

\clearpage
\section{Analytical results}\label{sec:analytic}
Before we address the emergence of dynamical attractors in numerical simulations of the various microscopic theories, it proves insightful to discuss (semi-)analytical solutions which can be obtained in the linear regime of weak metric perturbations $A \ll 1$.

Solutions for linearised metric perturbations can be obtained for hydrodynamics, kinetic theory and holography in a straightforward fashion. Specifically, we are interested in the response of the energy-momentum tensor $\delta T^{\mu\nu}$ to a metric perturbation $\delta g_{\alpha\beta}$, as described by the Green's function
\begin{align}\label{eq:linresponse}
    G^{\mu\nu,\alpha\beta}(x,y)=\frac{\delta T^{\mu\nu}(x)}{\delta g_{\alpha\beta}(y)}
\end{align}
where for the metric in \cref{eq:metric}, the perturbation is of the form $\delta g_{\eta\eta}(t)=2a(t)\delta a(t)$. Such equilibrium response functions have previously been studied in hydrodynamics, kinetic theory~\cite{Romatschke:2015gic,Kurkela:2017xis,Heller:2018qvh,Ochsenfeld:2023wxz,Gavassino:2024rck} and holography~\cite{Policastro:2002se,Kovtun:2005ev,Grozdanov:2019kge} to investigate the nature of hydrodynamic and non-hydrodynamic excitations, and we can easily adapt the calculations to the current problem. By going to Fourier space, where for an equilibrium system
\begin{align}
    G^{\mu\nu,\alpha\beta}(x,y)=\int \frac{d\omega}{2\pi} \int \frac{d^3k}{(2\pi)^3}~ G^{\mu\nu,\alpha\beta}(\omega,\vec{k})~e^{-i\omega (x^0-y^{0})} e^{+i \vec{k}(\vec{x}-\vec{y})}, \,
\end{align}
we are interested in the response of $\delta T^{tt}$ and $\delta T^{\eta\eta}$ to metric perturbations $\delta g_{\eta\eta}$, such that 
the relevant Green's functions for us are $G^{tt,\eta\eta}(\omega,k=0)
$ and $G^{\eta\eta,\eta\eta}(\omega,k=0)
$. Specifically, the linear response of the energy-momentum tensor is determined as 
\begin{align}
\label{eq:rtResp}
    \delta T^{\mu\nu}(t)=\int_{-\infty}^{+\infty}\frac{d\omega}{2\pi} G^{\mu\nu,\eta\eta}(\omega,k=0) ~\delta g_{\eta\eta}(\omega)~e^{-i\omega t}
\end{align}
where for metric perturbations as in \cref{eq:pert}
\begin{align}
    \delta g_{\eta\eta}(\omega)= \frac{1}{2} \left[ \frac{1}{\omega+\omega_0+i\epsilon} - \frac{1}{\omega-\omega_0+i\epsilon} \right]~2A\;,
\end{align}
in the linear regime ($A \ll 1$), where $\omega_0$ denotes the driving frequency $\omega$ of \cref{eq:pert} and we adopt the $i\epsilon$ prescription for the Heaviside theta function.

Based on \cref{eq:rtResp}  the linear response of the energy-momentum tensor in real time can then be formally evaluated by performing the integration using the residue theorem, taking into account that due to causality the retarded response Green's functions $G^{\mu\nu,\eta\eta}(\omega,k=0)$ as well as the perturbation $\delta g_{\eta\eta}(\omega)$ are holomorphic functions in the upper half plane. Specifically, if we assume that $G^{\mu\nu,\eta\eta}(\omega,k=0)$ only has a series of simple poles located at complex frequencies $\Omega_{n}$ with $\text{Im} \Omega_{n}<0$, we obtain that for $t>0$
\begin{align}
    \delta T^{\mu\nu}(t) &= \frac{i}{2}\left[G^{\mu\nu,\eta\eta}(\omega_0) - G^{\mu\nu,\eta\eta}(-\omega_0) \right] 2A \cos(\omega_0 t) \nonumber \\
    &\quad + \frac{1}{2} \left[G^{\mu\nu,\eta\eta}(\omega_0) + G^{\mu\nu,\eta\eta}(-\omega_0) \right] 2A \sin(\omega_0 t) \nonumber \\
    &{-} \sum_{n} 2A~\frac{i\omega_0}{\omega_0^2-\Omega_{n}^2}  \text{res}_{\Omega_{n}} (G^{\mu\nu,\eta\eta})~e^{-i \Omega_{n}t} \nonumber 
\end{align}
such that for a real-valued response function $G^{\mu\nu,\eta\eta}(x,y)$, we obtain
\begin{align}
    \label{eq:GeneralResponseLin}
   \delta T^{\mu\nu}(t)  &= -\text{Im}(G^{\mu\nu,\eta\eta}(\omega_0)) 2A \cos(\omega_0 t) + \text{Re}(G^{\mu\nu,\eta\eta}(\omega_0)) 2A \sin(\omega_0 t) \nonumber \\
    & - \sum_{n}  2A~\frac{i\omega_0}{\omega_0^2-\Omega_{n}^2}  \text{res}_{\Omega_{n}} (G^{\mu\nu,\eta\eta})~e^{-i \Omega_{n}t}\;.
\end{align} 
Based on the general result in \cref{eq:GeneralResponseLin}, we  obtain the linear response of the temperature $T(t)$ and shear stress $\pi(t)$ as 
\begin{align}
    \frac{T(t)}{T_0}&=1+\frac{\delta T^{tt}(t)}{4 T^{tt}_0}\;, \\
    \frac{\pi(t)}{P_{\rm eq}}&=\frac{\delta T^{\eta}_{\eta}(t)- \frac{1}{3} \delta T^{tt}(t)}{P_{\rm eq}}
\end{align}
Before we discuss the results for specific microscopic theories, it is important to point out that the structure of \cref{eq:GeneralResponseLin} already allows for a straightforward interpretation of the dynamical response. Namely, the terms in the first line describe the long-time response to the metric perturbation, which follows the harmonic driving frequency $\omega_0$, with a phase shift and amplitude determined by the behaviour of the real and imaginary parts of the Green's function $G^{\mu\nu,\eta\eta}(\omega_0)$ for real frequencies. Conversely, the terms in the second line are associated with the singularities of the retarded Green's function and describe the transient behaviour of the system before the emergence of a hydrodynamic attractor. While in \cref{eq:GeneralResponseLin} the amplitude of these transient excitations is solely determined by the frequency of the metric perturbation, more generally it may also depend on the initial conditions. However, causality requires $\text{Im}\, \Omega_{n}<0$, so that these transients decay exponentially, leading to the convergence to a hydrodynamic attractor.

\subsection{Navier--Stokes}
By considering the energy evolution \cref{eq:eom}, along with the constitutive relation \cref{eq:NSpi}, it is straightforward to determine the relevant Green's functions for the Navier--Stokes theory as
\begin{align}
    2G^{tt,\eta\eta}(\omega)&=-(e+P_{\text{eq}})\;, \\
    2G^{\eta\eta,\eta\eta}(\omega)&=-c_s^2(e+P_{\text{eq}}) + \frac{4}{3}\eta~i 
    \omega
\end{align}
where the first line follows directly from \cref{eq:eom}. Since $T^{\eta}_{~\eta}=P_{\text{eq}} + \pi$, the second line receives two distinct contributions, namely from the change of the equilibrium pressure, that is proportional to the speed of sound $c_s^2=1/3$, as well as from the shear stress. Evidently, the Green's functions are analytic, representing the fact that Navier--Stokes theory does not feature any transient 
excitations, such that in the linear regime temperature and shear stress directly follow the metric perturbation
\begin{align}
    \frac{T(t)}{T_0} &= 1-\frac{1}{3}A\sin \omega_0 t \;, \\
    \frac{\pi(t)}{P_{\text{eq}}} &= - \frac{16}{3} \frac{\eta T}{e+P_{\text{eq}}}  A \frac{\omega_0}{T_0} \cos(\omega_0 t)
\end{align}
where we used the conformal equation of state $e=3P_{\text{eq}}$ to express the shear stress in terms of the dimensionless ratio $\frac{\eta T}{e+P_{\text{eq}}}$. At vanishing chemical potential, $e+P_{\text{eq}}=Ts$, such that this ratio is equivalent to $\eta/s$ for all theories considered in this work except the scalar kinetic theory, where a finite chemical potential can develop.

Since in Navier--Stokes hydrodynamics the constitutive relation \cref{eq:NSpi} directly determines the shear stress as,
\begin{align}
    \frac{\pi}{P_{\text{eq}}}= - \frac{16}{3} \frac{\eta T}{e+P_{\text{eq}}} \frac{1}{T(t)} \left( \frac{\dot{a}}{a}\right)
\end{align}
it is also possible to obtain an explicit solution for the evolution of temperature in the non-linear regime, which takes the form
\begin{align}
     T(t)=\left(\frac{a(t=0)}{a(t)}\right)^{1/3} T_0 + \frac{4}{9} \frac{\eta T}{e+P_{\text{eq}}} \int_{0}^{t} dt' \left(\frac{a(t')}{a(t)}\right)^{1/3} \left(\frac{\dot{a}(t')}{a(t')}\right)^2
\end{align}
where the second term on the right-hand side describes the viscous heating, which is proportional to the square of the expansion rate, and therefore quadratic in the amplitude of the perturbation. When evaluated for a periodic perturbation, the temperature grows (on average) linearly with time, as can be seen explicitly by expanding the solution to second order in the amplitude $A$, where for $t>0$ the temperature evolution is given by
\begin{equation}
    T(t) = T_0\left(1-\frac{1}{3}A\sin \omega_0 t+\frac{2}{9}A^2 \sin^2 \omega_0 t\right) +\frac{4}{9}\frac{\eta T}{e+P_{\text{eq}}} A^2 \omega_0 \left(\frac{\omega_0 t}{2}+\frac{\sin 2\omega_0 t}{4}\right)+ \mathcal{O}(A^3)\label{eq:NS_temp}
\end{equation}
and the shear stress behaves as
\begin{equation}
    \frac{\pi}{P_{\text{eq}}}= - \frac{16}{3}\frac{\eta T}{e+P_{\text{eq}}} \frac{1}{T(t)} \frac{A \omega_0 \cos \omega_0 t}{1+A \sin \omega_0 t}\approx - \frac{16}{3}\frac{\eta T}{e+P_{\text{eq}}} A\frac{\omega_0 }{T_0} \left(\cos \omega_0 t- \frac{1}{3}A  \sin 2 \omega_0 t\right)+ \mathcal{O}(A^3)\label{eq:NS_pi}
\end{equation}

\subsection{Second-order hydrodynamics (MIS and rBRSSS)}
Following \cref{eq:eom} and \cref{eq:rbrsss_nice} the evolution equations for the temperature $T$ and normalised shear stress $\pi/P_{\text{eq}}$ in rBRSSS hydrodynamics can be recast as:
\begin{align}\label{Eq:MISeq1}
    \partial_t T&= - \frac{\dot{a}}{a} \left(\frac{1}{3} + \frac{1}{12}\frac{\pi}{P_{\text{eq}}}\right) T\;,\\ 
    C_{\tau_\pi}\, \partial_t \frac{\pi}{P_{\text{eq}}} + T \frac{\pi}{P_{\text{eq}}} + \frac{4}{3}\,C_{\tau_\pi}\frac{\dot{a}}{a}\frac{\pi}{P_{\text{eq}}} &= - \frac{16}{3}\,C_\eta \frac{\dot{a}}{a} + C_{\tau_\pi} \left(\frac{4}{3} + \frac{1}{3}  \frac{\pi}{P_{\text{eq}}} \right)  \frac{\dot{a}}{a}  \frac{\pi}{P_{\text{eq}}} \nonumber \\
    & \quad - \frac{8}{3}\,\frac{C_\kappa C_\eta}{T}\,\frac{\ddot{a}}{a} + \frac{T}{8}\,\frac{C_{\lambda_1}}{C_{\eta}}\left(\frac{\pi}{P_{\text{eq}}}\right)^2
\end{align}
where we used the conformal equation of state $e=3P_{\text{eq}}$ and expressed the transport coefficients in terms of the dimensionless quantities $C_\eta = \eta T/(e+P_{\text{eq}})$ and $C_{\tau_\pi} = T \tau_{\pi}$, $C_{\kappa}= T \kappa/\eta$, $C_{\lambda_1} = T\lambda_1/\eta$, which are constant in conformal theories but whose numerical values depend on the specific theory, see \cref{tab:coef}.
By linearising the evolution equations for small perturbations around equilibrium, it is then straightforward to obtain the relevant Green's functions as 
\begin{align}
\label{eq:MISResponse}
    2G^{tt,\eta\eta}(\omega)&=-(e+P_{\text{eq}})\;, \\
    2G^{\eta\eta,\eta\eta}(\omega)&=-c_s^2(e+P_{\text{eq}}) + \frac{4}{3}\eta \frac{i \omega + \kappa/(2\eta) \omega^2}{1-i \omega\tau_{\pi} }\label{eq:MISResponse1}
\end{align}
We first consider the case where $\kappa=0$, which is the case for all kinetic theories, as they do not exhibit a response to the space-time curvature~\cite{Moore:2012tc}. By performing the inverse Fourier transform according to \cref{eq:GeneralResponseLin}, the linearised solution for the evolution of the temperature and normalised shear stress for $t>0$ then takes the following form
\begin{align}
\label{eq:MISResponseT}
  \frac{T(t)}{T_0}&=1 -\frac{1}{3} A \sin(\omega_0 t)  \;, \\
\label{eq:MISResponsepi}
  \frac{\pi(t)}{P_{\rm eq}}&=-\frac{16}{3}  A\frac{\Big(\frac{\eta T}{e+P_{\text{eq}}}\Big)}{T \tau_\pi}~\frac{\omega_0 \tau_\pi}{1+\omega_0^2\tau_\pi^2}  \left[ \cos(\omega_0 t)+\omega_0 \tau_\pi \sin(\omega_0 t) - e^{-t/\tau_\pi}\right]\;.
 \end{align}
Here the last term in the square bracket, which corresponds to the contribution from the single pole of the Green's function, describes the transient behaviour, such that for times $t\gg \tau_\pi$ the solution settles onto an attractor curve, which, as discussed above, is determined by the Green's functions evaluated at the poles $\omega=\pm \omega_0 $ induced by the metric perturbation, and follows the oscillation of the metric with a frequency-dependent phase shift.

Since the space-time curvature $\sim \ddot{a}/a $ of the metric \cref{eq:metric} exhibits a singular behaviour at initial time $t=0$, including the curvature response $\kappa$ leads to a discontinuous behaviour of the shear stress, where according to the rBRSSS evolution equations
\begin{align}
    \pi(0^{+})-
    \pi(0^{-}) = -
    \frac{2}{3} \frac{\kappa}{\tau_\pi}
     \left[\frac{\dot{a}}{a}(0^{+}) - \frac{\dot{a}}{a}(0^{-}) \right]\;.
\end{align}
Since we are primarily interested in the dynamics following the metric quench, as opposed to the quench itself, we can always choose the initial conditions such that the system is in equilibrium right after the quench, i.e.\ $\pi(0^{+})=0$, in which case the hydrodynamic response is given by
\begin{align}  
  \frac{\pi(t)}{P_{\rm eq}}=&-\frac{16}{3}  A\frac{\Big(\frac{\eta T}{e+P_{\text{eq}}}\Big)}{T \tau_\pi}~\frac{\omega_0 \tau_\pi}{1+\omega_0^2\tau_\pi^2}  \left[ \left(1+\frac{1}{2}\frac{(\omega_0 \tau_\pi)^2}{T\tau_\pi}\left(\frac{T\kappa}{\eta}\right)\right)\cos(\omega_0 t) \right.\notag\\
  &\left. +\left(\omega_0 \tau_\pi - \frac{1}{2}\frac{\omega_0 \tau_\pi}{T \tau_\pi}\left(\frac{T \kappa}{\eta}\right)\right) \sin(\omega_0 t) - \left(1+\frac{1}{2}\frac{(\omega_0 \tau_\pi)^2}{T\tau_\pi}\left(\frac{T\kappa}{\eta}\right)\right) e^{-t/\tau_\pi}\right]\label{eq:lin_MIS_pi}
\end{align}
\subsection{RTA kinetic theory}\label{sec:RTA_analytic}

Notably, the relevant Green's functions of the energy-momentum tensor can also be calculated analytically in kinetic theory in the relaxation-time approximation. We follow~\cite{Romatschke:2015gic,Ochsenfeld:2023wxz} and consider a linearised perturbation of the phase-space distribution function, which follows the frequency $\omega$ of the metric perturbation according to
\begin{align}
    \delta f = \delta f_{\omega}e^{-i\omega t}
\end{align}
By linearising the Boltzmann equation in the framework of RTA, we can solve for the perturbation $\delta f_{\omega}$
\begin{align}
    \delta f_{\omega } = \frac{1}{1-i\omega \tau_R}\left(-\frac{p}{T}\delta T_{\omega}-i\omega\tau_R\frac{p_{\|}^2}{p}\frac{\delta g_{\omega}}{2}\right)\frac{\partial f_{\text{eq}}}{\partial p}
\end{align}
where $\delta g_\omega$ denotes the amplitude of the metric perturbation, i.e.\ $\delta g_{\eta\eta}=\delta g_{\omega} e^{-i\omega t} =2\delta a_{\omega} e^{-i\omega t}$ ($\delta a_\omega$ refers to the amplitude of the scale factor perturbation) 
and $\delta T_{\omega}$ is the change in equilibrium temperature associated with $\delta f_{\omega}$, which needs to be self-consistently determined from the Landau matching condition. By calculating the linear response of the energy-momentum tensor, we obtain 
\begin{align}
    \frac{\delta T^{tt}_\omega}{T_0^{tt}}= \frac{\nu \int \frac{d^3p}{(2\pi)^3} p \delta f_{\omega} } {T_0^{tt}} = \frac{4\left(\frac{\delta T_\omega}{T_0}+i\omega\tau_R\frac{\delta g_{\omega}}{6}\right)}{1-i\omega\tau_R} 
\end{align}
such that upon enforcing the Landau matching condition $\frac{4\delta T_\omega}{T_0} = \frac{\delta T^{tt}_\omega}{T_{0}^{tt}}$, we obtain a closed equation for the temperature perturbation $\delta T_\omega$ depending linearly on the metric perturbation $\delta g_\omega$. By re-inserting the solution for $
\delta T_{\omega}$ into $\delta f_{\omega}$, we can then immediately determine the 
Green's functions
$G^{\mu\nu,\eta\eta}(\omega)=\delta T_\omega^{\mu\nu}/\delta g_{\omega}$ 
for the energy-momentum tensor components $\delta T^{tt}$ and $\delta T^{\eta\eta}$ as
\begin{align}
    2G^{tt,\eta\eta}(\omega)&=-(e+P_{\text{eq}}) \\
    2G^{\eta \eta,\eta\eta}(\omega)&=-c_s^2(e+P_{\text{eq}})+\frac{4}{15}(e+P_{\text{eq}})\frac{i\omega\tau_R}{1-i\omega\tau_R}
\end{align}
Strikingly, upon identifying the transport coefficients
\begin{align*}
    \eta = \frac{\tau_R}{5}(e+P_{\text{eq}})\; \qquad
    \tau_{\pi} = \tau_R    
\end{align*}
the Green's functions in kinetic theory in the relaxation-time approximation exactly match the results of MIS hydrodynamics in \cref{eq:MISResponse} for $\kappa=0$, indicating that MIS hydrodynamics and RTA kinetic theory behave identically in the linear regime.\footnote{Note that this agreement only holds if the metric perturbations are spatially homogeneous $(k=0)$, and we refer to~\cite{Romatschke:2015gic} for a more general discussion of Green's functions of the energy-momentum tensor in kinetic theory.}

Beyond the linear regime, the solution to the Boltzmann equation in the relaxation-time approximation can be formally expressed through the implicit relation (cf.~\cite{Du:2021fok}),
\begin{align}
   & f(t,p_T, p^\eta) =  e^{-\int_0^t \frac{dt'}{\tau_R(t')}} f_\text{eq}(p_T, p^\eta a^2(t), T_0)+\int_{0}^t \frac{dt'}{\tau_R(t')}  e^{-\int_{t'}^t \frac{dt''}{\tau_R(t'')}} f_\text{eq}\left(p_T,p^\eta \frac{a^2(t)}{a^2(t')}, T(t')\right),
\end{align}
where for convenience we use the transverse momentum $p_T=\sqrt{p_x^2+p_y^2}$ and the coordinate momentum $p^\eta=p_\|/a(t)$ as independent variables, and the temperature $T(t)$ is determined self-consistently as the instantaneous equilibrium temperature after Landau matching
\begin{align}
e(t)&=\nu\int\frac{d^2 p_T\, dp^\eta}{(2\pi)^3}\,a\, \sqrt{p_T^2+a^2(p^\eta)^2}\,  f(t,p_T, p^\eta)\equiv \nu \frac{\pi^2}{30} T^4 
\end{align}
By performing the momentum integral on both sides, we can thus obtain an implicit integral equation for the temperature evolution
\begin{align}
   & T^4(t) =  e^{-\int_0^t \frac{dt'}{\tau_R(t')}} T_0^4 I(a(t))+\int_{0}^t \frac{dt'}{\tau_R(t')}  e^{-\int_{t'}^t \frac{dt''}{\tau_R(t'')}} T(t')^4 I\left( \frac{a(t)}{a(t')}\right),
\end{align}
where the function $I(x)$ is given by
\begin{align}
  I(x>1)&=  \frac{1}{2} \left( \frac{1}{x^2} + \frac{\tan^{-1}\sqrt{x^2-1}}{\sqrt{x^2-1}}\right)\\
I(x<1)&=  \frac{1}{2} \left( \frac{1}{x^2} + \frac{\tanh^{-1}\sqrt{1-x^2}}{\sqrt{1-x^2}}\right)\\
I(x=1)&=1
\end{align}

\subsection{Free streaming}
We note that in the absence of interactions, i.e.\ for a free-streaming system, the above function readily provides the exact evolution of the energy density, where for $t>0$
\begin{align}
\frac{T(t)}{T_0} &=I^{1/4}(a(t))
\end{align}
and the shear stress for a free-streaming evolution can be written  as
\begin{align}
\frac{\pi(t)}{P_{\text{eq}}} &= 3\,\frac{a^2(t)\,I(a(t))-1}{a^2(t)\,(a^2(t)-1)\,I(a(t))}-1 \label{eq:FS_pi}
\end{align}
which is regular at $a=1$, where $\pi=0$, such that for small amplitudes ($|a(t)-1| \ll 1$) one has $\frac{\pi(t)}{P_{\rm eq}} \approx -\frac{16}{15} (a(t)-1)$.
Specifically, in the linear regime, the free-streaming Green's functions of the energy-momentum tensor are given by
\begin{align}
    2G^{tt,\eta\eta}(\omega)&=-(e+P_{\text{eq}}) \\
    2G^{\eta\eta,\eta\eta}(\omega)&=-c_s^2(e+P_{\text{eq}})-\frac{4}{15}(e+P_{\text{eq}})
\end{align}
which coincides with the limit $\tau_R \rightarrow \infty$ of the Green's functions in the relaxation-time approximation.

\subsection{Holography}\label{sec:holography_analytic}

In the holographic description, the linear response of the energy-momentum tensor to a metric perturbation can be determined from the retarded stress-tensor correlator, which is obtained by studying linearised gravitational fluctuations around the AdS$_5$ black brane dual to the thermal equilibrium state of the boundary theory at temperature $T$. In ingoing Eddington--Finkelstein coordinates this background is the static case of \cref{Eq:HoloMetric}, with $\mathcal{S} = 1/r$, $\mathcal{P} = 0$ and
$\mathcal{A} = (1 - r^4/r_h^4)/r^2$, where $r_h = 1/(\pi T)$ is the horizon. The
fluctuations obey the Einstein equations \labelcref{Eq:Einstein} linearised around this background, and regularity at the future horizon, equivalent to the ingoing condition in these coordinates, selects the retarded solution \cite{Son:2002sd}.

We compute the retarded correlator of the stress tensor \cite{Policastro:2002se,Policastro:2002tn,Kovtun:2005ev}, and use its dynamical part, together with the appropriate contact terms, to construct the response function $G^{\mu\nu,\eta\eta}$ appearing in \cref{eq:linresponse} (see \cref{app:holoappendix}).
Owing to the translation invariance of the boundary theory, the gravitational fluctuation
$h_{MN}$ admits a Fourier decomposition,
\begin{align} \label{eq:holopertubation}
    h_{MN}(r,t,\vec{x}) = \int \frac{d\omega\, d^3k}{(2\pi)^4}\,
    e^{-i(\omega t - \vec{k}\cdot\vec{x})}\, h_{MN}(r,\omega,\vec{k}),
\end{align}
with frequency $\omega$ and momentum $\vec{k}$. For generic momentum, $h_{MN}$ splits into
the tensor (helicity-2), shear (helicity-1) and sound (helicity-0) channels according to
their transformation properties under the residual rotational symmetry about the momentum direction. Each sector has its own gauge-invariant master field, giving rise to distinct retarded correlators. Since
the drive is spatially homogeneous, we restrict to $k=0$, where there is no distinguished momentum direction and all channels obey the same second-order differential equation \labelcref{Eq:gauge-invariant equation} for the traceless gauge-invariant combination $Z = r^2\left(2h_{\eta\eta} - h_{xx} - h_{yy}\right)$. In terms of the dimensionless radial coordinate $r \to r/r_h$, which places the horizon at $r=1$, it reads
\begin{align} \label{Eq:gauge-invariant equation}
    Z''(r) - \frac{\left(4 - f(r) - 4 i \mathfrak{w}\, r\right) Z'(r)}{r f(r)}
    - \frac{6 i \mathfrak{w}\, Z(r)}{r f(r)} = 0,
\end{align}
with $f(r) = 1 - r^4$ and the dimensionless frequency $\mathfrak{w} = \omega/(2\pi T)$. We solve \cref{Eq:gauge-invariant equation} with regularity imposed at the horizon, which selects the retarded solution, and obtain the retarded correlator by expanding this solution towards the boundary and extracting its near-boundary coefficients. The dynamical part of the retarded
correlator is then
\begin{align} \label{eq:GR_dyn}
    G_R^{\eta\eta,\eta\eta}(\mathfrak{w})\Big|_{\rm dyn}
    = -\frac{2}{3}\pi^2 N_c^2 T^4\, \frac{R(\mathfrak{w})}{Q(\mathfrak{w})},
\end{align}
in the convention of Ref.~\cite{Kovtun:2005ev}, where at low frequency
$R/Q = i\mathfrak{w}/2 + \mathcal{O}(\mathfrak{w}^2)$ yields the familiar
$G_R \simeq -\tfrac{4}{3} i\omega\eta$. Here $Q(\mathfrak{w})$ is the coefficient of the non-normalisable mode and therefore specifies the boundary source, while $R(\mathfrak{w})$ corresponds to the coefficient of the normalisable mode after subtracting the logarithmic piece as discussed in \cref{app:holoappendix}. Combining this dynamical correlator with the necessary contact terms and using the metric perturbation relation $\delta g_{\eta\eta} = 2 \delta a$, we obtain the response function of \cref{eq:linresponse},
\begin{align}
    2G^{tt,\eta\eta}(\mathfrak{w}) &= -(e + P_\text{eq}),
    \label{Eq:Gtt_holo} \\
    2G^{\eta\eta,\eta\eta}(\mathfrak{w}) &= -c_s^2 (e + P_\text{eq}) -G_R^{\eta\eta,\eta\eta}(\mathfrak{w})\Big|_{\rm dyn} + \frac{7}{18}\pi^2 N_c^2 T^4\, \mathfrak{w}^4.
    \label{Eq:Longitudinal_StressCorrelator}
\end{align}
The frequency-independent terms coincide with those of Navier--Stokes, MIS and RTA, and cancel in the combination $2 \left(G^{\eta\eta,\eta\eta} - c_s^2 G^{tt,\eta\eta}\right)$.

At low frequency, the response reduces to second-order hydrodynamics,
\begin{align} \label{Eq:BRSSS_GR}
   2 \left( G^{\eta\eta,\eta\eta} - c_s^2 G^{tt,\eta\eta} \right)
    = \frac{4}{3}\, i\omega\eta \left( 1 + i\tau_\pi\omega - i\frac{\kappa}{2\eta}\,\omega
    + \mathcal{O}(\omega^2) \right),
\end{align}
where the values of $\eta$, $\tau_\pi$ and $\kappa$ for strongly coupled $\mathcal{N} = 4$ supersymmetric Yang--Mills theory are
\begin{align} \label{Eq:SYM_transport}
    \eta = \frac{\pi}{8} N_c^2 T^3, \qquad
    \tau_\pi = \frac{2 - \log 2}{2\pi T}, \qquad
    \kappa = \frac{\eta}{\pi T},
\end{align}
in agreement with the MIS response \cref{eq:MISResponse1} expanded to the same order. 
Beyond the hydrodynamic regime, quasinormal modes occur at frequencies for which the non-normalisable coefficient vanishes,
\begin{align} \label{Eq:QNMcondition}
    Q(\mathfrak{w}_n) = 0.
\end{align}
At these frequencies, the bulk solution is ingoing at the horizon and normalisable at the boundary, with no external source.
The lowest quasinormal frequencies are $  \mathfrak{w}_1 =\pm1.55973-1.37334\,i,~  \mathfrak{w}_2 =\pm2.58476-2.38179\,i, $
while all higher poles lie deeper in the lower half of the complex frequency plane (see \cref{Fig:QNM spectrum}).
Since $R/Q$ enters the dynamical part of the response, these quasinormal frequencies are poles of both the retarded correlator and the corresponding response function~\cite{Kovtun:2005ev}, thereby establishing the correspondence between gravitational quasinormal modes and the non-hydrodynamic excitations of the boundary theory.

\subsection{Approach to the attractor across theories}
\label{sec:Sec3Disc}
\begin{figure}[t]
    \centering
        \includegraphics[width=0.9\linewidth]{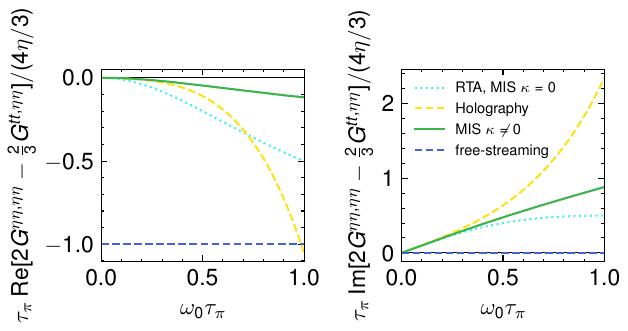}
    \caption{Normalised linear shear-stress response on the attractor, $\tau_\pi\big[2G^{\eta\eta,\eta\eta}-\tfrac{2}{3}G^{tt,\eta\eta}\big]/(4\eta/3)$, versus driving frequency $\omega_0\tau_\pi$ (left: real part; right: imaginary part) for holography (yellow dashed), RTA kinetic theory and MIS with $\kappa=0$, which coincide (cyan dotted), MIS with the holographic $\kappa\neq0$ (green solid) and free streaming (dashed). Navier--Stokes corresponds to $\mathrm{Re}=0$, $\mathrm{Im}=\omega_0\tau_\pi$, which all theories approach for $\omega_0\tau_\pi\ll1$.
    }
    \label{Fig:Compare_HBRsss}
\end{figure}

Now that we have established the retarded correlation functions for the different dynamical theories under consideration, we can investigate the linear response of the system to the metric perturbation in \cref{eq:metric}. Based on \cref{eq:GeneralResponseLin} the long-time behaviour of the shear stress $\pi$ is characterised by an attractor, where the magnitude of the shear stress is determined by the real and imaginary parts of $2 \left(G^{\eta\eta,\eta\eta}-\frac{1}{3}G^{tt,\eta\eta}\right)$ evaluated at the (real) frequency $\omega_0$ of the metric perturbation. We analyse the properties of this attractor in \cref{Fig:Compare_HBRsss}, where we present the results for the real and imaginary parts of the normalised shear-stress response, as a function of $\omega_0\tau_{\pi}$, i.e.\ the frequency $\omega_0$ of the metric perturbation in units of the shear relaxation time $\tau_{\pi}$. The normalisation $\tau_{\pi} \Big[2G^{\eta\eta,\eta\eta}-\frac{2}{3}G^{tt,\eta\eta}\Big]/(4\eta/3)$ is chosen in order to obtain universal hydrodynamic predictions for the Navier--Stokes and MIS theories. While at small frequencies $\omega_0 \tau_\pi \ll 1$ all theories converge towards the Navier--Stokes behaviour, large discrepancies emerge at larger frequencies. While the response in MIS hydrodynamics with $\kappa=0$ and RTA kinetic theory remains identical for all frequencies (cf.~\cref{eq:MISResponse}), and eventually approaches the free-streaming behaviour at asymptotically large frequencies $\omega_0 \tau_\pi \gg 1$, already for $\omega_0 \tau_\pi \sim 1$ the holographic attractor shows a much larger response compared to the hydrodynamic/kinetic theory attractor.

\begin{figure}[H]
    \centering
    \includegraphics[width=0.45\linewidth]{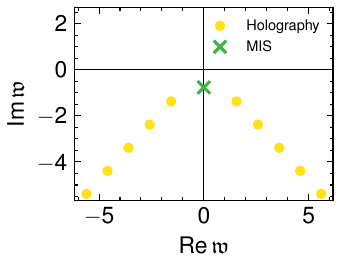}
    \includegraphics[width=0.47\linewidth]{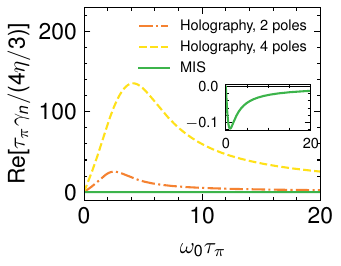}
    \caption{Left: non-hydrodynamic poles of the shear-stress response in the complex $\mathfrak{w}=\omega/(2\pi T)$ plane: the lowest holographic quasinormal-mode pairs $\pm\mathrm{Re}\,\mathfrak{w}_n+i\,\mathrm{Im}\,\mathfrak{w}_n$ (yellow dots) and the purely damped MIS/RTA pole $\mathfrak{w}_\text{MIS}=-i/(2\pi T\tau_\pi)$ (green cross). Right: residue-weighted excitation amplitude $\mathrm{Re}[\tau_\pi\gamma_n/(4\eta/3)]$ versus driving frequency $\omega_0\tau_\pi$, summed over the lowest one (2 poles, orange dash-dotted) and two (4 poles, yellow dashed) holographic QNM pairs, compared to MIS with $\kappa\neq0$ (green solid), which is small and negative (see inset).}
    \label{Fig:QNM spectrum}
\end{figure}

Differences in the analytic structure of the retarded Green's functions manifest themselves in the transient behaviour of the shear-stress response, which characterises the approach to the hydrodynamic attractor following the quench of the metric in \cref{eq:metric}. In accordance with \cref{eq:GeneralResponseLin}, the transient behaviour is associated with the singularities of the retarded Green's functions, where for MIS hydrodynamics and RTA kinetic theory, there is a single isolated pole\footnote{Note that for finite $k\neq 0$ the analytic structure of the retarded Green's function in kinetic theory is more complicated, see e.g.\ \cite{Romatschke:2015gic,Kurkela:2017xis,Ochsenfeld:2023wxz}.} located at $\Omega=-i/\tau_{\pi}$, which leads to an exponential approach to the hydrodynamic attractor on time scales $\sim \tau_{\pi}$. The contribution of this mode is given by \footnote{\cref{eq:GeneralResponseLin} assumes equilibrium
before the quench, $\pi(0^-)=0$, as in the standard
equilibrium retarded-correlator treatment in
holography~\cite{Kovtun:2005ev}. If instead
$\pi(0^+)=0$ is imposed, as in \cref{eq:lin_MIS_pi}
and \cref{sec:numerics}, the factor
$(1-\kappa/(2\eta\tau_\pi))$ in the MIS transient
amplitude is replaced by
$\bigl(1+(\omega_0\tau_\pi)^2
\kappa/(2\eta\tau_\pi)\bigr)$.}
\begin{align}  \label{eq:residueMIS}
    \gamma_{\rm MIS}(\omega_0)=\frac{i\omega_0}{\omega_0^2-\Omega^2}  \text{res}_{\Omega} \Big( 2G^{\eta\eta,\eta\eta}- \frac{2}{3}G^{tt,\eta\eta}\Big)= - \frac{4}{3} \eta \frac{\omega_0 \left( 1- \kappa/(2 \eta \tau_\pi)\right)}{\omega_0^2\tau_\pi^2+1} 
\end{align}
Conversely, the holographic theory possesses an infinite tower of non-hydrodynamic modes, located at discrete frequencies $\Omega_{n} \equiv  2 \pi T \mathfrak{w}_n $ with finite real and imaginary parts as illustrated in \cref{Fig:QNM spectrum}.  Since the non-hydrodynamic modes appear in complex-conjugate pairs, the approach towards the late-time attractor is characterised by a superposition of exponentially decaying oscillations, whose individual contribution to the shear stress is determined by 
\begin{align}
    \gamma_{n}(\omega_0)=\frac{i\omega_0}{\omega_0^2-\Omega_{n}^2}  \text{res}_{\Omega_{n}} \Big( 2G^{\eta\eta,\eta\eta}- \frac{2}{3}G^{tt,\eta\eta}\Big)= i
     \frac{\omega_0} {\omega_0^2-\Omega_n^2} \rho_n(\Omega_n)  
\end{align}
where the residue function $\rho_n(\Omega_n)$ comes from the dynamical part of the correlator $G^{\eta\eta,\eta\eta}$ which may be represented as, 
\begin{align} \label{Eq:Holo_GR}
   \rho_n(\Omega_n) \equiv \text{res}_{\Omega_{n}} \Big( 2G^{\eta\eta,\eta\eta}- \frac{2}{3}G^{tt,\eta\eta}\Big)
    =  \frac{2}{3}\pi^2 N_c^2 T^4 
    \frac{R(\Omega_n)}
   {Q'(\Omega_n)}  .
\end{align}

We compare the excitation of non-hydrodynamic modes in \cref{Fig:QNM spectrum}, where we show the normalised amplitude $\tau_{\pi} \gamma / (4 \eta/3)$ as a function of $\omega_0 \tau_{\pi}$ for the first four non-hydrodynamic modes of the holographic theory, and compare them to MIS hydrodynamics/RTA kinetic theory. 
We observe that for the lowest holographic QNM pair, the excitation amplitude is largest around $\omega_0\tau_\pi \simeq 2.5$, indicating that this pair is excited most strongly to the driven transient response at intermediate frequencies. Including the second pair makes a substantial contribution to the transient excitation amplitudes, and the amplitude further increases for higher QNM pairs, indicating the importance of higher QNMs within the chosen range of driving frequencies\footnote{Since the
amplitudes grow with $n$, the sum over modes shown in \cref{Fig:QNM spectrum}
does not converge. For any $t>0$, however, the stronger damping of the higher
modes renders the transient response finite.}. In contrast, MIS theory contains only a single purely damped relaxation pole. Its residue-weighted excitation amplitude is much smaller in magnitude than the holographic contribution shown in the figure, resulting in a weaker transient response.

\clearpage
\section{Numerical results}\label{sec:numerics}

Now that we have established a (semi-)analytical picture of the linear response in frequency space, in this section we 
can compare the temperature and shear-stress tensor evolution in different microscopic and macroscopic theories undergoing a periodic expansion and contraction given by \cref{eq:pert}.

\subsection{Small-amplitude limit \texorpdfstring{$A\ll 1$}{A << 1}}\label{sec:small_amplitude}
We begin by considering small-amplitude oscillations with amplitude $A$ and varying driving
frequency $\omega$, where the full numerical solutions can be directly compared to the strictly linear solutions derived in \cref{sec:analytic}. The ratio of shear viscosity to entropy density is fixed to $\eta/s = 0.56$, the value obtained in pure-glue QCD for $\lambda=g^2N_c=10$~\cite{Arnold:2003zc,Boguslavski:2024kbd} (see \cref{tab:coef}), 
across all kinetic and hydrodynamic descriptions to enable a fair comparison, whereas in holography $\eta/s=1/(4\pi)$. When presenting results for the free-streaming case, 
we adopt the relaxation-time approximation (RTA) values of $\eta/s$ and $\tau_\pi$ wherever 
these quantities appear.

\subsubsection{Macroscopic theories}\label{sec:macroscopic}
\begin{figure}
    \centering
    \includegraphics[width=\linewidth]{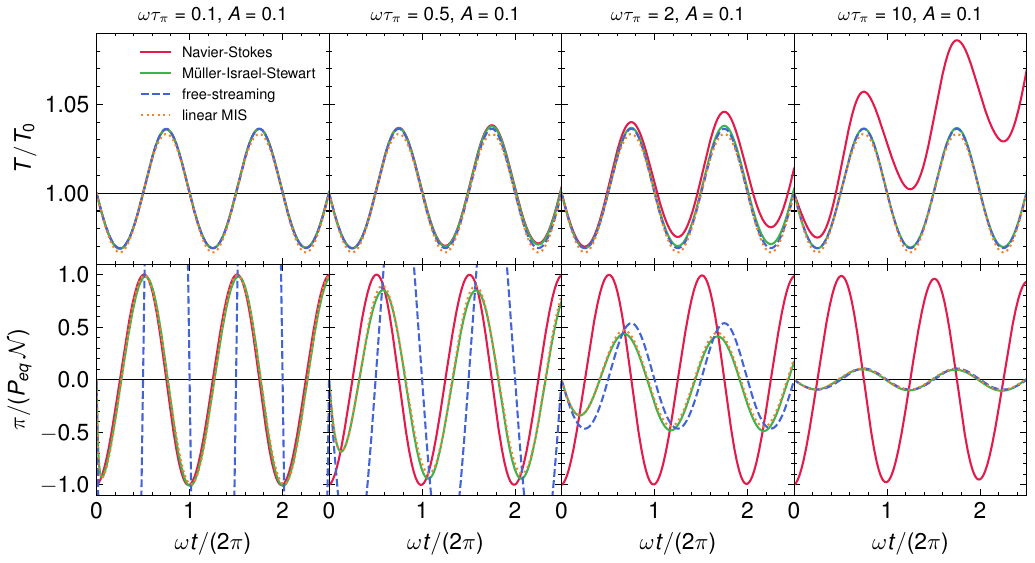}
    \caption{Response to a periodic drive with small amplitude $A = 0.1$ across 
    representative frequencies spanning the range $\omega\tau_\pi \ll 1$ to 
    $\omega\tau_\pi \gg 1$ for different macroscopic theories. \textit{Upper panels:} Relative temperature $T/T_0$ as a function 
    of time. \textit{Lower panels:} Normalised shear-stress tensor $\pi/(P_\mathrm{eq}\mathcal{N})$.}
    \label{fig:MIS}
\end{figure}
In the limit of small frequencies, $\omega \tau_\pi \ll 1$ (measured in units of the inverse 
relaxation time $\tau_\pi^{-1}$), all descriptions converge to the universal hydrodynamic 
limit, where the system responds quasi-statically to the external drive. In \cref{fig:MIS} we show the relative temperature $T/T_0$ (upper panel) and the normalised 
shear-stress tensor $\pi/(P_\mathrm{eq}\mathcal{N})$  (lower panel) as functions of time, $\omega t$. The normalisation $\mathcal{N}=\tfrac{16}{3} A\omega\,(\eta/s)/T_0$ is chosen such that the Navier--Stokes result for small frequency lies in the range $[-1,1]$, see \cref{eq:NS_pi}.
In \cref{fig:MIS} we include results for
Navier--Stokes (NS, solid red line) and Müller--Israel--Stewart (MIS, solid green line) 
hydrodynamics. For comparison, we also include the free-streaming (FS) result (blue dashed line) 
and the linearised MIS solution (orange dotted line) in \cref{eq:MISResponseT,eq:MISResponsepi}.

When considering low driving frequencies ($\omega\tau_\pi \ll 1$), the temperature evolution (upper left panel) is identical 
across all descriptions. This universality arises because, at linear order in the amplitude 
$A$, viscous heating is absent and the temperature evolves isentropically, as reflected in the universal Green's function $G^{tt,\eta\eta}$ found in \cref{sec:analytic}.
In this regime, NS, MIS and FS are indistinguishable at the level of the temperature.
However, comparing to the strictly linear-order MIS expansion (orange dotted line), we see some small deviations. These are caused by the first quadratic 
corrections in \cref{eq:NS_temp}, which arise simply from the non-linearity of the expansion rate $\dot a/a$.

At high driving frequencies ($\omega\tau_\pi \gg 1$), however, the quadratic terms in 
\cref{eq:NS_temp} become significant (upper right panel) and the temperature in the Navier--Stokes description grows due 
to viscous heating. By contrast, the MIS solution does not exhibit this behaviour 
and instead converges towards the free-streaming result for large driving frequencies $\omega\tau_\pi \gg 1$. The suppression of heating in MIS can be 
traced to the behaviour of the shear-stress tensor $\pi$ (lower panels), which is responsible for entropy 
production: unlike the instantaneous Navier--Stokes constitutive relation, the MIS framework 
introduces a finite relaxation time $\tau_\pi$ that prevents $\pi$ from tracking the 
rapidly oscillating velocity gradients, cf.\ the linear MIS solution \cref{eq:MISResponsepi}.
This is clearly visible in the lower panels of \cref{fig:MIS}. At low frequencies, the 
Navier--Stokes and MIS shear stresses show much smaller deviations from equilibrium than 
free streaming, since interactions act to isotropise the energy-momentum tensor. However, as the 
frequency increases, the instantaneous Navier--Stokes response grows and eventually 
exceeds the free-streaming amplitude. Since a realistic system with a finite microscopic relaxation 
time $\tau_\pi$ cannot follow the rapid oscillations of the background, the shear-stress 
anisotropy saturates at the non-interacting (free-streaming) limit for 
$\omega\tau_\pi \gg 1$. While this saturation is correctly captured by the MIS framework, it is missed entirely by the Navier--Stokes description, which has no built-in memory of the relaxation 
timescale. We finally note a characteristic $\pi/2$ phase offset between the Navier--Stokes and 
free-streaming solutions for $\pi/P_\mathrm{eq}$. This phase difference reflects the 
distinct physical mechanisms at work: in Navier--Stokes theory the shear stress responds 
instantaneously and in phase with the velocity gradient $\propto {\dot a}/{a}$, while in free streaming the pressure anisotropy is proportional to the current value of $a$, see \cref{eq:FS_pi}.

\begin{figure}
    \centering
    \includegraphics[width=\linewidth]{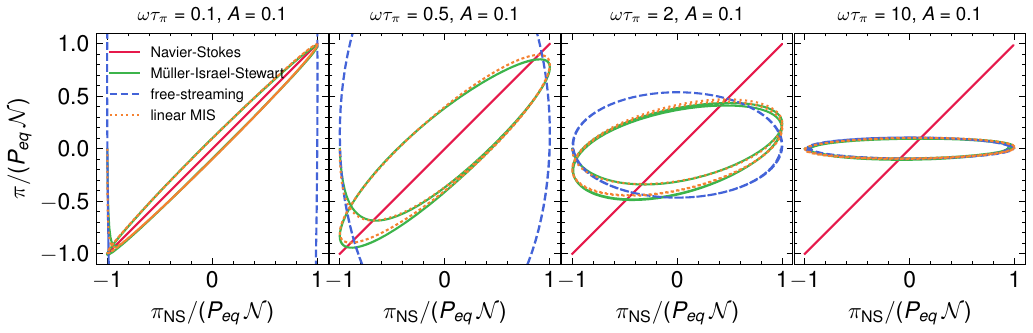}
    \caption{Normalised shear stress for different macroscopic theories plotted versus the Navier--Stokes expectation. The linearised MIS solution is plotted against the linearised NS expectation, while all other theories are plotted against the full non-linear NS solution.}
    \label{fig:MIS2}
\end{figure}

Next, to highlight the emergence of a dynamical attractor, we present in \cref{fig:MIS2} a parametric plot of the normalised shear-stress tensor shown in \cref{fig:MIS} (lower panel) versus the Navier--Stokes expectation. One observes that for $\omega \tau_\pi\ll 1$, after a short transient time, the MIS theory forms a narrow ellipse aligned with the NS diagonal, where the width of the ellipse is given by $\omega \tau_\pi$. For increasing frequency the ellipse widens and tilts. Since the early-time transient is
governed by the relaxation time $\tau_\pi$, relaxation becomes slower relative to the larger drive frequency (cf.\ the $e^{-t/\tau_\pi}$ term in \cref{eq:MISResponsepi}), until eventually for $\omega \tau_\pi \gg1$ the pressure response aligns with the free-streaming result and the relaxation time no longer plays a role. 

The cyclic attractor is well captured by the linear MIS theory, \cref{eq:MISResponsepi}, which after the decay of transients asymptotes to an ellipse given by
\begin{equation}
\left(\frac{1+(\omega \tau_\pi)^2}{\omega \tau_\pi}y- \frac{1}{\omega \tau_\pi}x\right)^2 +x^2 = 1 \label{eq:ellipse_asymptote} \end{equation}
where $x$ and $y$ are the normalised shear stresses $\pi/(P_\text{eq} \mathcal{N})$ for NS and MIS consistently linearised to leading order in $A$.

\subsubsection{Kinetic theories}

Having established the behaviour of the macroscopic theories, we now turn to the
microscopic kinetic descriptions. We compare the three kinetic theories introduced in
\cref{sec:kinetic}: the relaxation-time approximation (RTA), scalar $\phi^4$
theory, and Yang--Mills effective kinetic theory (EKT). As in the macroscopic
case, we fix the ratio of shear viscosity to entropy density to $\eta/s = 0.56$
across all the kinetic theories. As a macroscopic reference, we additionally show the
MIS result. We stress that, since the second-order coefficient
$C_{\tau_\pi}/C_\eta$ differs between the theories (see \cref{tab:coef}),
fixing $\eta/s$ implies a slightly different relaxation time $\tau_\pi$, and
hence a slightly different physical frequency $\omega$, for a given value of
$\omega\tau_\pi$. As before, the initial state is the equilibrium one with
$\pi(t=0)=0$.

\begin{figure}
    \centering
    \includegraphics[width=\linewidth]{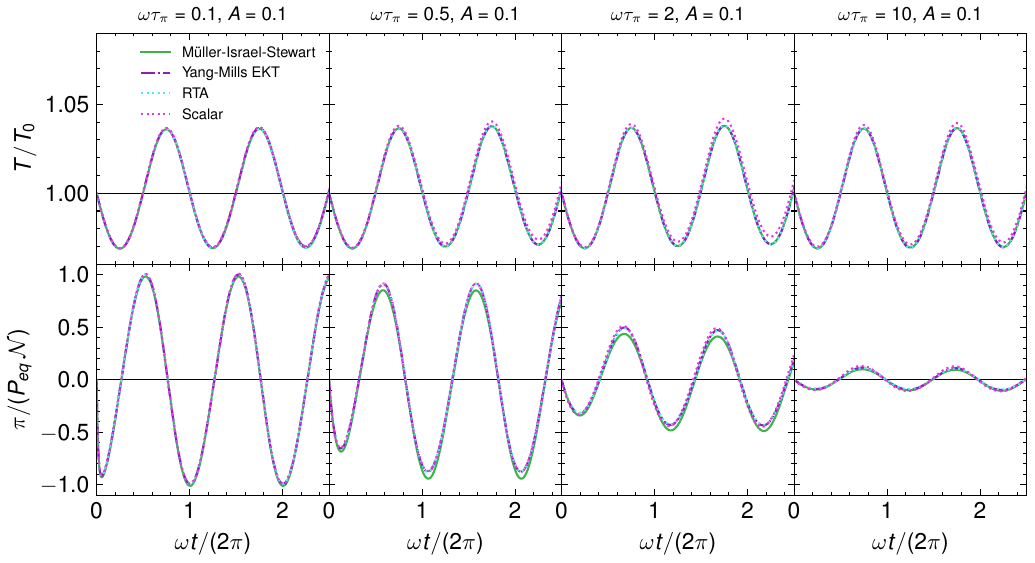}
    \caption{Kinetic theory response to a periodic drive of small amplitude
    $A=0.1$, for representative frequencies spanning
    $\omega\tau_\pi \ll 1$ to $\omega\tau_\pi \gg 1$. \textit{Upper panels:}
    relative temperature $T/T_0$ as a function of time. \textit{Lower panels:}
    normalised shear stress $\pi/(P_\mathrm{eq}\mathcal{N})$ as a function of
    time. Results are shown for the RTA,
    scalar and Yang--Mills kinetic theories, together with the MIS reference. The
    normalisation $\mathcal{N}=\tfrac{16}{3}A\omega\,(\eta/s)/T_0$ is as in
    \cref{fig:MIS}.}
    \label{fig:kinetic}
\end{figure}

In \cref{fig:kinetic} we show the temperature evolution $T/T_0$ (upper panels)
and the normalised shear stress $\pi/(P_\mathrm{eq}\mathcal{N})$ as a function of
time (lower panels) for the same representative frequencies as in
\cref{fig:MIS}. The most striking feature is the near-universal agreement between
the three kinetic theories. Only at larger frequencies does the scalar kinetic theory show a slight difference in the evolution of the normalised shear stress and temperature, which we attribute to the fact that the scalar kinetic theory is particle-number conserving and develops a non-zero chemical potential in response to the metric perturbation.

\begin{figure}
    \centering
    \includegraphics[width=\linewidth]{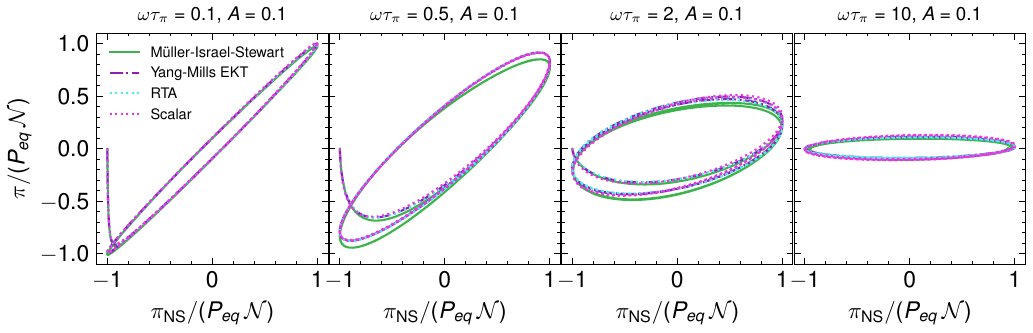}
    \caption{Parametric trajectory of the normalised shear stress
    $\pi/(P_\mathrm{eq}\mathcal{N})$ against its Navier--Stokes expectation
    $\pi_\mathrm{NS}/(P_\mathrm{eq}\mathcal{N})$, for a periodic drive of small
    amplitude $A=0.1$ and the same representative frequencies as in
    \cref{fig:kinetic}. Results are shown for the RTA,
    scalar and Yang--Mills kinetic theories, together with the MIS reference. The
    normalisation $\mathcal{N}=\tfrac{16}{3}A\omega\,(\eta/s)/T_0$ is as in
    \cref{fig:MIS}.}
    \label{fig:kinetic_ps}
\end{figure}

Strikingly, the kinetic theory results also track the macroscopic MIS reference closely, and
reproduce the same frequency dependence discussed in \cref{sec:macroscopic}. Slight differences in the normalised shear stress $\pi/(P_\mathrm{eq}\mathcal{N})$
nonetheless persist at intermediate frequencies, and are most pronounced in the vicinity of the minima and maxima of the shear stress. Since the MIS theory agrees with RTA kinetic theory at the strictly linear level (see \cref{sec:RTA_analytic}), this suggests that even for $A=0.1$ there are non-linear effects, which are not captured exactly by MIS hydrodynamics. 

The emergence of a dynamical attractor is made more transparent in the parametric trajectories of
\cref{fig:kinetic_ps}, which should be compared with the macroscopic result in
\cref{fig:MIS2}. After a short transient, governed by $\tau_\pi$, each theory
settles onto a closed cyclic attractor. For $\omega\tau_\pi \ll 1$ this attractor
is a narrow ellipse aligned with the Navier--Stokes diagonal, whose width is set
by $\omega\tau_\pi$. As the frequency increases, the ellipse widens and tilts away
from the diagonal, reflecting the growing phase lag between $\pi$ and the drive.
Remarkably, the trajectories of all three kinetic theories collapse onto one
another and onto the MIS curve, with only a small spread developing at larger frequencies.
This demonstrates that, at small amplitude, the late-time cyclic attractor is
universal across weakly coupled kinetic theories and is faithfully described by
the much simpler MIS theory.

\subsubsection{Holography}
\begin{figure}
    \centering
    \includegraphics[width=\linewidth]{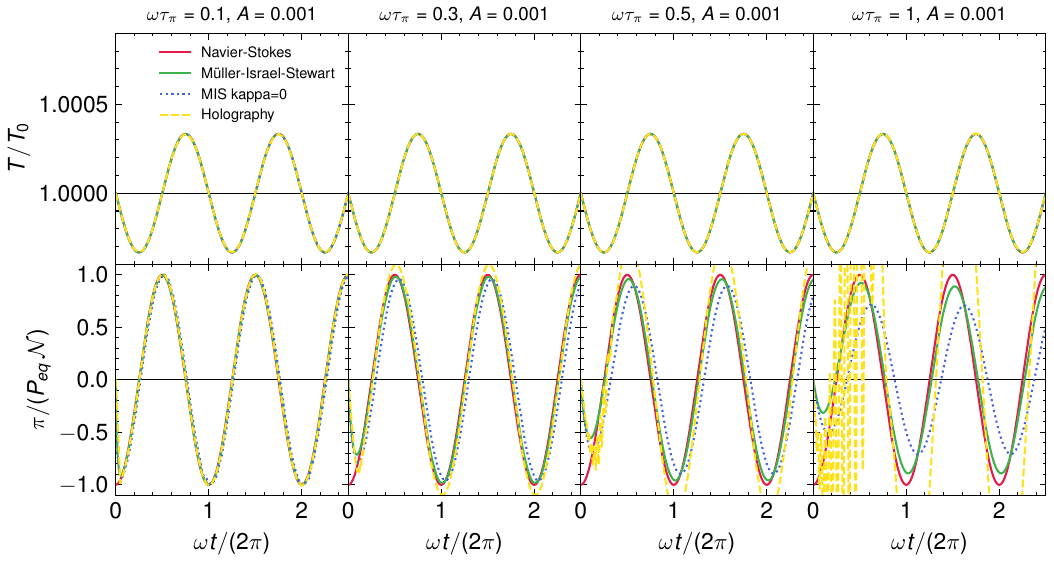}
    \caption{Holographic response to a periodic drive of small amplitude
    $A=0.001$, for frequencies $\omega\tau_\pi = 0.1$, $0.3$, $0.5$ and $1$. The holographic simulation is initialised at $t_0 = 0$ with $ A_4(t_0=0) = -1$.
    \textit{Upper panels:} relative temperature $T/T_0$ as a function of time.
    \textit{Lower panels:} normalised shear stress $\pi/(P_\mathrm{eq}\mathcal{N})$
    as a function of time. The holographic result is compared with Navier--Stokes,
    MIS, and MIS with the curvature coupling switched off ($\kappa = 0$). The
    normalisation $\mathcal{N}=\tfrac{16}{3}A\omega\,(\eta/s)/T_0$ is as in
    \cref{fig:MIS}.}
    \label{fig:holo}
\end{figure}

Finally, in \cref{fig:holo} we show results for strongly coupled holographic simulations with $\eta/s=1/(4\pi)$. Due to numerical limitations, we work at the
even smaller amplitude $A=0.001$ and lower frequencies $\omega\tau_\pi=0.1,\,0.3,\,0.5,\,1.0$, and compare
holography against Navier--Stokes and MIS hydrodynamics. Since holography is the
only theory considered here with a non-vanishing second-order curvature coupling,
$C_\kappa/C_\eta = 4$ (see \cref{tab:coef}), we show both the full MIS result and the
MIS result with $\kappa=0$ to isolate its effect.

\begin{figure}
    \centering
    \includegraphics[width=0.5\linewidth]{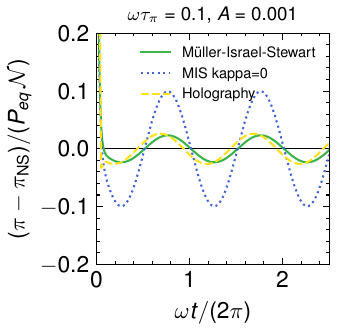}
    \caption{Difference of the normalised shear stress $\pi/(P_\mathrm{eq}\mathcal{N})$
    from the Navier--Stokes expectation $\pi_\mathrm{NS}/(P_\mathrm{eq}\mathcal{N})$ as a
    function of time, for a periodic drive of small amplitude $A=0.001$ at the lowest
    frequency $\omega\tau_\pi = 0.1$. The holographic result is compared with MIS and MIS
    with the curvature coupling switched off ($\kappa = 0$). The normalisation
    $\mathcal{N}=\tfrac{16}{3}A\omega\,(\eta/s)/T_0$ is as in \cref{fig:MIS}.}
    \label{fig:holot_Tsub}
\end{figure}

The temperature evolution shown in the upper panels of \cref{fig:holo} is indistinguishable across all theories and frequencies. This is expected, since $G^{tt,\eta\eta}$ in \cref{Eq:Gtt_holo} coincides with that of all other theories. For the normalised shear-stress tensor (lower panels), we find good agreement with NS and MIS at low frequencies. At larger frequencies, however, discrepancies in the oscillation amplitude emerge: holography yields a larger amplitude than NS, while both MIS variants fall
below NS, with MIS $\kappa=0$ deviating furthest from holography. 
This mirrors the linear-response comparison in \cref{Fig:Compare_HBRsss}, where the holographic $\text{Im}\,G$ grows much faster with $\omega\tau_\pi$ than in MIS. Clearly, this is in contrast to the kinetic theories, where agreement with MIS hydrodynamics persists across all frequencies studied. We further scrutinise this behaviour in \cref{fig:holot_Tsub}, where we show the residual $[\pi - \pi_\mathrm{NS}]/(P_\mathrm{eq}\mathcal{N})$ at the lowest frequency $\omega\tau_\pi=0.1$. This comparison reveals that the deviations from Navier--Stokes are largest not at the peaks but at the steep zero crossings. Crucially, MIS with $\kappa \neq 0$ closely tracks the holographic result, while the $\kappa=0$ prediction deviates significantly, demonstrating the importance of the curvature coupling for an accurate description of the holographic response, in line with \cref{eq:lin_MIS_pi,Eq:BRSSS_GR}.

\begin{figure}
    \centering
    \includegraphics[width=\linewidth]{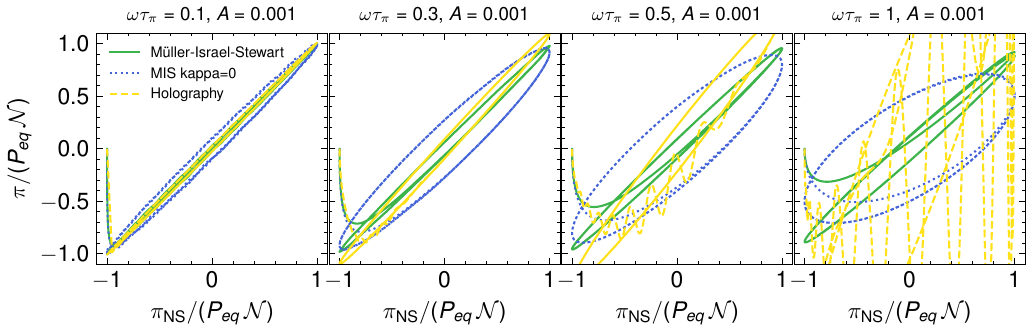}
    \caption{Parametric trajectory of the normalised shear stress
    $\pi/(P_\mathrm{eq}\mathcal{N})$ against its Navier--Stokes expectation
    $\pi_\mathrm{NS}/(P_\mathrm{eq}\mathcal{N})$, for a periodic drive of small
    amplitude $A=0.001$ and the same frequencies as in \cref{fig:holo}. The
    holographic result is compared with MIS, and MIS with the
    curvature coupling switched off ($\kappa = 0$). The normalisation
    $\mathcal{N}=\tfrac{16}{3}A\omega\,(\eta/s)/T_0$ is as in \cref{fig:MIS}.}
    \label{fig:holo_ps}
\end{figure}

A qualitatively new feature of the holographic response is the rapid ringing in the initial transient. This can also be observed in the phase-space portrait shown in \cref{fig:holo_ps}, where for the two largest frequencies the holographic
curve clearly rings around the cyclic attractor. In accordance with our discussion in \cref{sec:Sec3Disc} we can attribute these oscillations to the excitation of the non-hydrodynamic quasinormal modes, whose complex frequencies have a non-vanishing real part
and therefore produce genuine ringing, and whose excitation amplitude grows with $\omega\tau_\pi$ (\cref{Fig:QNM spectrum}).
This is in sharp contrast to
kinetic theory, whose non-hydrodynamic sector only features singularities on the imaginary frequency axis, resulting in an exponential decay towards the cyclic attractor. \\

\subsection{Non-linear response in RTA kinetic theory}\label{sec:nonlinear}
\begin{figure}
    \centering
    \includegraphics[width=\linewidth]{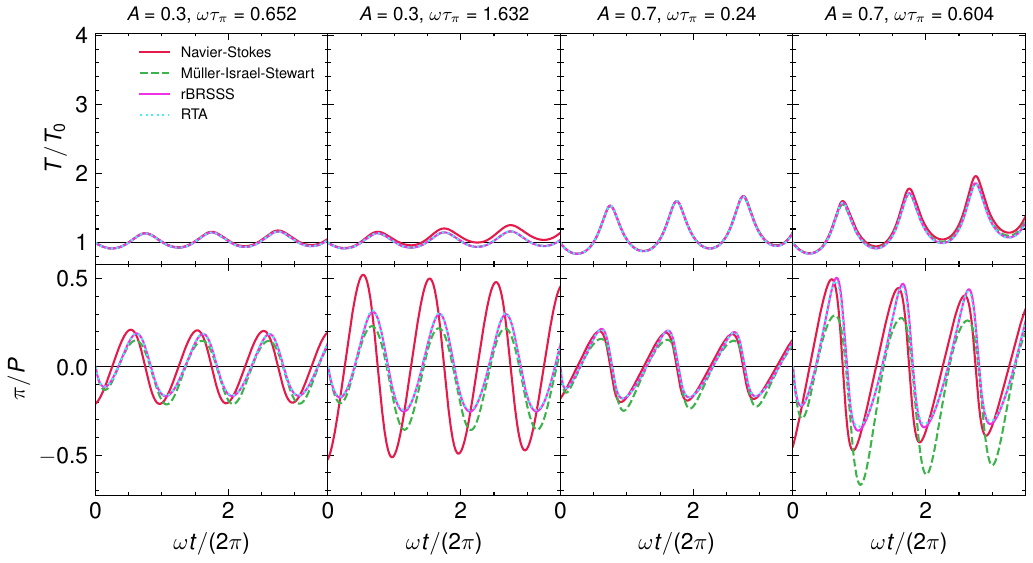}
    \caption{Response to a periodic drive in the non-linear (large-amplitude) regime,
    for amplitudes $A=0.3$ and $A=0.7$ at the frequencies indicated above each column.
    \textit{Upper panels:} relative temperature $T/T_0$ as a function of time.
    \textit{Lower panels:} shear stress $\pi/P_\mathrm{eq}$ as a function of time. In each
    panel we show Navier--Stokes (solid red line), Müller--Israel--Stewart (MIS, dashed green
    line), rBRSSS with its non-linear second-order terms (solid magenta line), and RTA kinetic theory
    (dotted cyan line).}
    \label{fig:nonlin_time}
\end{figure}

\begin{figure}
    \centering
    \subfigure[RTA, MIS and Navier--Stokes]{\includegraphics[width=0.49\linewidth]{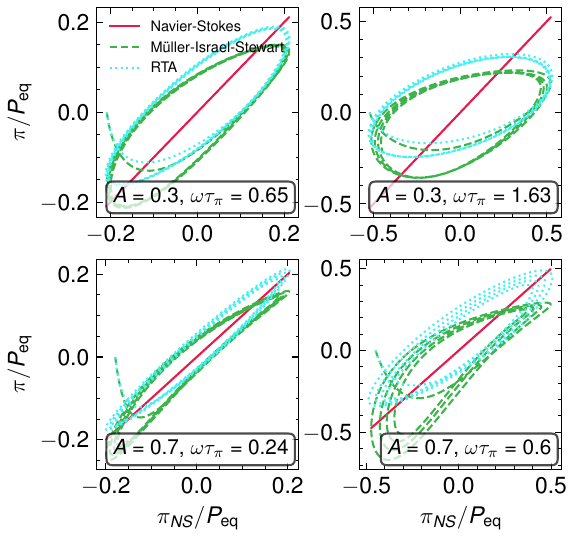}}
    \subfigure[RTA, rBRSSS and Navier--Stokes]{\includegraphics[width=0.49\linewidth]{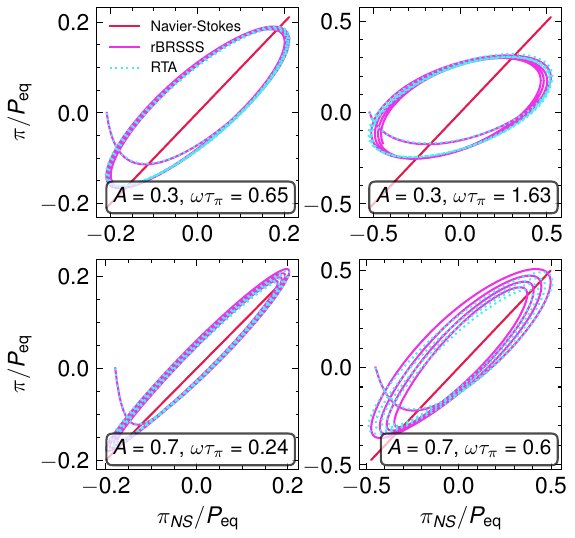}}
    \caption{Parametric trajectory of the shear stress $\pi/P_\mathrm{eq}$ against its
    Navier--Stokes expectation $\pi_\mathrm{NS}/P_\mathrm{eq}$ in the non-linear regime, for
    the amplitudes and frequencies labelled in each panel. \textit{(a)} RTA kinetic theory
    (dotted cyan line) compared with MIS (dashed green line) and Navier--Stokes (solid red
    line). \textit{(b)} RTA kinetic theory (dotted cyan line) compared with rBRSSS with its
    non-linear second-order terms (solid magenta line) and Navier--Stokes (solid red line).}
    \label{fig:nonlin_ps}
\end{figure}

Next, we investigate the non-linear response that emerges for large-amplitude perturbations. We will limit ourselves to RTA kinetic theory
and compare it with MIS and rBRSSS hydrodynamics. In general, hydrodynamics is expected to
apply when the Knudsen number is small~\cite{Niemi:2014wta},
\begin{equation}
    \text{Kn} \equiv |\tau_{\pi} \theta|
.\end{equation}
For small-amplitude driving, the maximum Knudsen number reached in RTA kinetic theory is
$\text{Kn}_\text{max}=  A\omega \tau_\pi=\frac{15}{16}\mathcal{N}$, where $\mathcal N$ is the amplitude of $\pi/P_\text{eq}$ for Navier--Stokes theory.
For large-amplitude drives we instead use the empirical formula
\begin{align}
    \text{Kn}_\text{max}= \max\left(\left|\frac{5 \eta}{s T_0} \cdot \frac{T_0}{T}\cdot\frac{\dot{a}}{a}\right|\right) \approx \frac{5 \eta}{sT_0} \frac{A \omega \sqrt{1 - \frac{4}{9} A^2}}{(1 - \frac{2}{3} A^2) \cdot (1 + \frac{2}{9} A^2)}.\label{eq:Kn_max} 
\end{align}

In \cref{fig:nonlin_time,fig:nonlin_ps} we demonstrate the importance of the non-linear
second-order terms of rBRSSS in the limit of large amplitude and Knudsen number.
\Cref{fig:nonlin_time} shows the time evolution of the relative temperature $T/T_0$ (upper
panels) and the shear stress $\pi/P_\mathrm{eq}$ (lower panels) for amplitudes $A=0.3$ and
$A=0.7$ at fixed $\text{Kn}_\text{max}\approx 0.2$ and $0.5$. These correspond to
frequencies $\omega\tau_\pi=0.65,\,1.63$ for $A=0.3$ and
$\omega\tau_\pi=0.24,\,0.60$ for $A=0.7$.
In each case the RTA result is compared with
Navier--Stokes, MIS, and rBRSSS with its non-linear second-order terms.
As the amplitude grows, the non-linear terms become important: the temperature evolution
departs from the simple sinusoidal oscillations of the linear response and acquires a growing
viscous-heating contribution. The Navier--Stokes solution overestimates this temperature
increase at larger Knudsen number, whereas RTA, MIS, and rBRSSS agree very well up to large
amplitude and Knudsen number. The non-linear second-order terms of rBRSSS are therefore not
needed to capture the temperature evolution.

The shear stress (lower panels of \cref{fig:nonlin_time}) tells a different story. At larger
amplitude and Knudsen number, MIS and rBRSSS visibly disagree, with rBRSSS tracking the RTA
solution more closely, since here the non-linear second-order terms become relevant. This is seen
most clearly in the phase-space portraits of \cref{fig:nonlin_ps}, which compare the stress
evolution against MIS (left panel) and rBRSSS (right panel). The rBRSSS trajectory follows RTA
far more faithfully than MIS, confirming that the non-linear second-order terms of rBRSSS provide a significantly
better description of the stress evolution.

\begin{figure}
    \centering
    \includegraphics[width=0.9\linewidth]{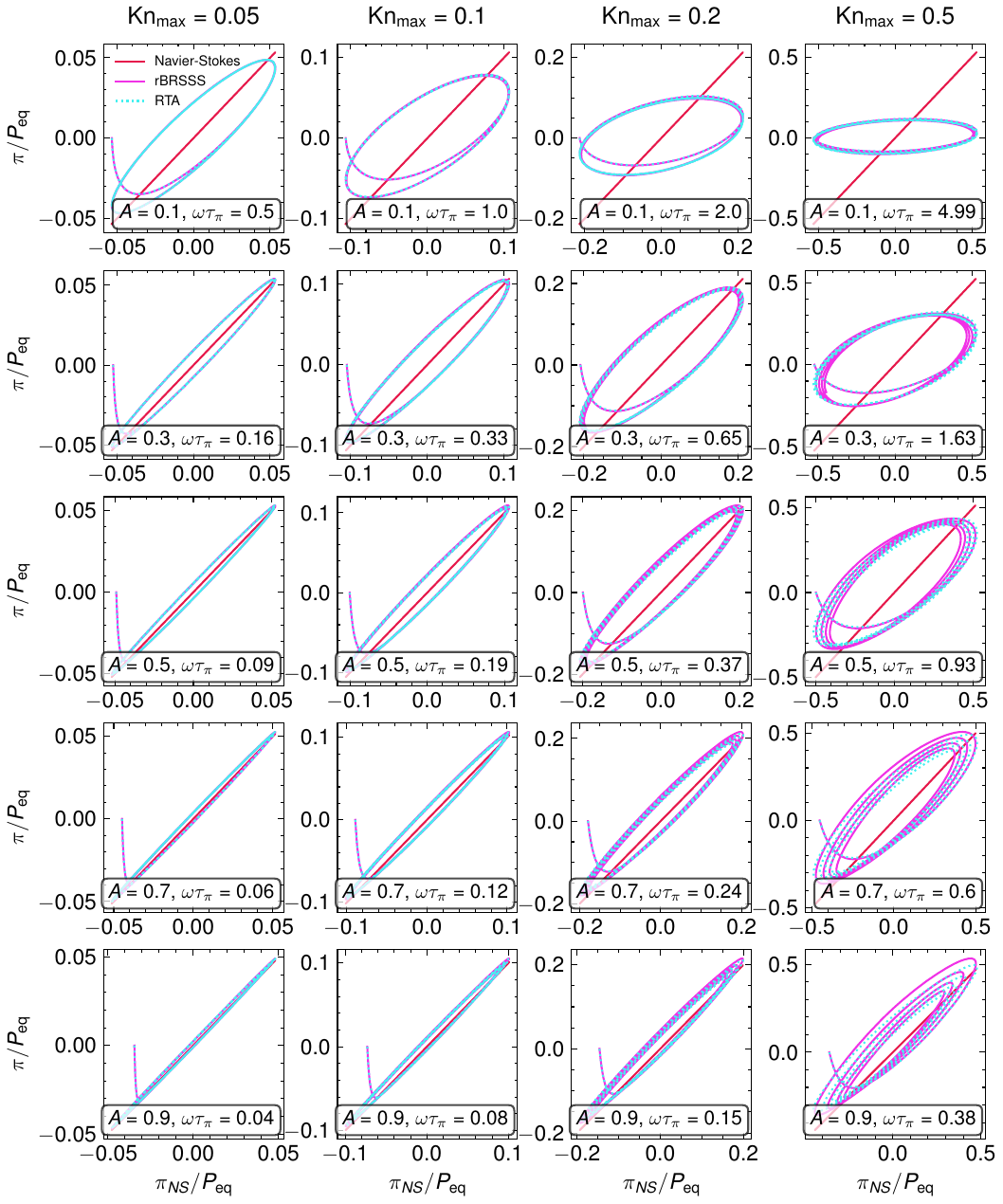}
    \caption{Scan of the non-linear regime for amplitudes $A=\{0.1,\,0.3,\,0.5,\,0.7,\,0.9\}$
    (from top to bottom) and maximum Knudsen number $\text{Kn}_\text{max}=\{0.05,\,0.1,\,0.2,\,0.5\}$
    (from left to right). Each panel shows the parametric trajectory of the shear stress
    $\pi/P_\mathrm{eq}$ against its Navier--Stokes expectation $\pi_\mathrm{NS}/P_\mathrm{eq}$
    for Navier--Stokes (solid red line), rBRSSS with its non-linear second-order terms (solid magenta line),
    and RTA kinetic theory (dotted cyan line).}
    \label{fig:nonlin_scan}
\end{figure}

Finally, \cref{fig:nonlin_scan} presents a systematic scan of the evolution over amplitudes
$A=\{0.1,\,0.3,\,0.5,\,0.7,\,0.9\}$ (rows) and maximum Knudsen numbers
$\text{Kn}_\text{max}=\{0.05,\,0.1,\,0.2,\,0.5\}$ (columns), with the frequency adjusted for
each pair according to \cref{eq:Kn_max}. We compare the RTA, rBRSSS, and Navier--Stokes
trajectories.
As long as one of the two parameters remains small, the agreement between RTA and rBRSSS
persists almost independently of the other, as seen in the top row ($A=0.1$) and the left
column ($\text{Kn}_\text{max}=0.05$). For small amplitude this is expected, since at linear order rBRSSS reduces to MIS, which coincides with RTA (\cref{sec:RTA_analytic}). The discrepancy grows only once both parameters become
large, with rBRSSS then tending to underestimate the stress relative to RTA. We also observe a
deformation of the rBRSSS ellipse that is controlled predominantly by the amplitude rather
than by the Knudsen number: the larger the amplitude, the stronger the bending. The RTA
trajectory bends as well, but less pronouncedly and in the opposite direction. Since the
frequency sets the phase shift for both RTA and rBRSSS, the spirals have a comparable width
along the diagonals running from top left to bottom right, where the frequencies are of
similar order. Moreover, because the average temperature rises more rapidly at larger
amplitude, the spiralling towards the asymptotic ellipse is correspondingly accelerated, so
that the trajectories collapse onto the attractor more quickly as the amplitude increases --
again, this holds for both RTA and rBRSSS.

\subsection{Long-time behaviour in RTA kinetic theory}\label{sec:longtime}

\begin{figure}
    \centering
    \includegraphics[width=0.9\linewidth]{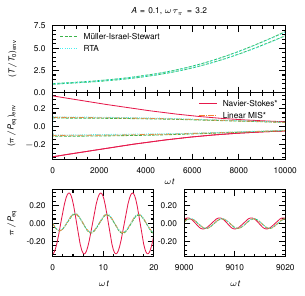}
    \caption{Long-time response to a small-amplitude periodic drive ($A=0.1$, $\omega\tau_\pi=3.2$) in hydrodynamics and RTA kinetic theory. \textit{Upper panel:} temperature envelope $T_{\text{env}}$ as a function of $\omega t$, showing the steady viscous heating of the medium. \textit{Middle panel:} envelope of the long-time shear-stress evolution. \textit{Lower panels:} normalised shear stress $\pi/P_\mathrm{eq}$ at early times (left) and at late times (right). Results are shown for Navier--Stokes ($\ast$), RTA, full (numerical) MIS, and linear MIS ($\ast$). The star indicates that the temperature entering the analytic expressions is promoted from a constant to the dynamical MIS temperature of the upper panel.}
    \label{fig:LargeTimesEvol}
\end{figure}

We finally turn to the long-time fate of the system under a sustained periodic drive.
For numerical convenience we work within RTA kinetic theory and MIS hydrodynamics, which can be integrated to
very long times at modest cost. We take a small-amplitude, high-frequency drive, $A=0.1$ and
$\omega\tau_\pi=3.2$, for which the response initially deviates strongly from the
Navier--Stokes expectation, but is well described within MIS hydrodynamics.
Under a sustained drive, viscous heating steadily injects energy into the medium, so that its
average temperature grows over many cycles. This heating is displayed in the upper panel of
\cref{fig:LargeTimesEvol}, where we plot the temperature envelope $T_\text{env}$, obtained by numerically extracting the minimum and maximum of the temperature over one oscillation cycle. 

Since the relaxation time decreases with rising temperature as $\tau_{\pi} \sim 1/T$, the effective drive
strength $\omega\tau_\pi$ decreases as a function of time, and we therefore expect the system to relax towards the
Navier--Stokes regime at asymptotically late times.
The middle panel  of
\cref{fig:LargeTimesEvol} shows the envelope of the shear-stress-to-equilibrium-pressure ratio,
comparing the full RTA evolution with the instantaneous linear Navier--Stokes and linear MIS
predictions. The latter two are obtained by inserting the instantaneous temperature of the
upper panel into \cref{eq:NS_pi,eq:lin_MIS_pi}. Since they are not solutions of the corresponding
equations of motion, we mark them with an asterisk. At early times the Navier--Stokes envelope
overshoots the true response, as expected at large $\omega\tau_\pi$, but the discrepancy shrinks
as the temperature rises. In contrast, the linear MIS predictor tracks the shear-stress
amplitude throughout the evolution. The two lower panels, which resolve the oscillations at early
(left) and late (right) times, show that its phase is reproduced equally well: the RTA kinetic evolution, full MIS
evolution and the linear MIS$^{*}$ curve overlap almost perfectly. The long-time shear-stress
response is thus captured by \cref{eq:lin_MIS_pi} once the slow temperature drift is accounted
for, and at late times even Navier--Stokes provides an accurate description.

\begin{figure}
    \centering
    \includegraphics[width=0.75\linewidth]{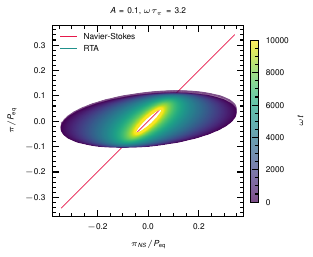}
    \caption{Parametric trajectory of the RTA shear stress $\pi/P_\mathrm{eq}$ against its Navier--Stokes expectation $\pi_\mathrm{NS}/P_\mathrm{eq}$ for a small-amplitude drive ($A=0.1$, initial $\omega\tau_\pi=3.2$), with colour encoding the elapsed time $\omega t$. The straight line marks the Navier--Stokes diagonal $\pi=\pi_\mathrm{NS}$. As the medium heats up and $\omega\tau_\pi$ decreases, the trajectory spirals inward, from a wide ellipse tilted away from the diagonal onto a thin ellipse aligned with it, illustrating the approach of RTA to Navier--Stokes at late times.}
    \label{fig:LargeTimes}
\end{figure}

This convergence is seen most clearly in the phase-space portrait of \cref{fig:LargeTimes}, where the RTA shear stress is plotted parametrically against its Navier--Stokes expectation with colour encoding the elapsed time. As the medium heats and $\omega\tau_\pi$ decreases, the cyclic attractor spirals inward: both the amplitude of the shear stress and its phase lag relative to Navier--Stokes diminish, so that the wide, tilted ellipse of the early evolution collapses onto the narrow ellipse aligned with the Navier--Stokes diagonal that characterises the hydrodynamic limit. This is precisely the behaviour encoded in the asymptotic ellipse \cref{eq:ellipse_asymptote}, whose width and tilt both vanish as $\omega\tau_\pi \to 0$.

\section{Conclusions}\label{sec:conclusions}

In this work we have presented a comprehensive study of hydrodynamic attractors
in conformal systems subject to a periodic expansion and contraction in one
spatial dimension. In contrast to the monotonic Bjorken expansion, where the
system relaxes towards equilibrium and the Navier--Stokes constitutive relation,
the periodic drive continuously excites the shear sector and the system instead
settles onto a \emph{cyclic attractor} -- a closed trajectory in the space of
the shear stress $\pi$ and its Navier--Stokes expectation $\pi_\mathrm{NS}$. By
employing hydrodynamic (Navier--Stokes, MIS and rBRSSS), weakly coupled kinetic
(RTA, scalar $\phi^4$ and Yang--Mills EKT) and strongly coupled holographic
descriptions, introduced in \cref{sec:setup}, we have mapped out the response of the energy-momentum tensor as a
function of the drive amplitude $A$ and the dimensionless frequency
$\omega\tau_\pi$.

Our starting point in \cref{sec:analytic} was a (semi-)analytic linear-response analysis of the driven
system, which provides a transparent framework for interpreting the subsequent
numerical results. Since the metric perturbation in \cref{eq:pert} is
spatially homogeneous, the response of the energy-momentum tensor is entirely
determined by the retarded correlators $G^{tt,\eta\eta}(\omega,k=0)$ and
$G^{\eta\eta,\eta\eta}(\omega,k=0)$, which we computed for Navier--Stokes, MIS,
RTA kinetic theory, free streaming and holography. Evaluating the real-time
response by contour integration, \cref{eq:GeneralResponseLin}, shows that the
dynamics separates into two pieces with a clear physical meaning: a steady
contribution that oscillates at the driving frequency $\omega_0$, whose amplitude
and phase shift are fixed by the real and imaginary parts of the Green's function
at the \emph{real} frequency $\omega_0$, and a transient contribution built from
the singularities of the retarded Green's function in the lower half of the complex frequency plane. The
former is precisely the cyclic attractor, the latter the memory of the initial
condition, which decays as a function of time. The cyclic attractor is therefore an
equilibrium property of the theory, encoded in the equilibrium response function,
while the approach to it is controlled by the non-hydrodynamic spectrum.

A key analytic result of \cref{sec:analytic} is that at vanishing wavenumber the RTA Green's functions
coincide \emph{exactly} with those of MIS hydrodynamics, once the transport
coefficients are identified as $\eta = \tau_R(e+P_{\rm eq})/5$ and
$\tau_\pi = \tau_R$. This identity explains the universality of the
small-amplitude attractor observed numerically, and it also fixes the two limiting
behaviours: for $\omega_0\tau_\pi \ll 1$ the response reduces to Navier--Stokes,
whereas for $\omega_0\tau_\pi \gg 1$ it saturates at the free-streaming value,
recovered as the $\tau_R \to \infty$ limit of the RTA correlators. By contrast,
the holographic attractor, shown in \cref{Fig:Compare_HBRsss}, departs from
the hydrodynamic/kinetic one already at $\omega_0\tau_\pi \sim 1$ and exhibits a
considerably larger response, reflecting the structure of the strongly
coupled theory beyond second-order hydrodynamics.
The
transient sectors differ just as sharply: MIS and RTA possess a single purely
damped singularity at $\Omega = -i/\tau_\pi$, giving a monotonic exponential
approach to the attractor, while holography carries an infinite tower of
quasinormal modes with non-vanishing real parts, so that the approach proceeds
through damped oscillations. Weighting each pole by its residue and by the
kinematic factor $i\omega_0/(\omega_0^2-\Omega_n^2)$, \cref{Fig:QNM spectrum}
shows that the lowest quasinormal pair is excited most strongly around
$\omega_0\tau_\pi \simeq 2.5$, that the second pair still contributes
significantly in the frequency range of interest, and that all holographic
amplitudes exceed the single MIS residue by a large margin, which is the analytic origin
of the pronounced ringing seen in the numerical holographic transients.

In \cref{sec:small_amplitude}, we complemented these analytic predictions with full numerical
simulations at small drive amplitudes, including results for the
scalar and Yang--Mills kinetic theories.
In the small-amplitude regime, the response is organised by the single parameter
$\omega\tau_\pi$. For slow driving, $\omega\tau_\pi \ll 1$, all descriptions
converge to a universal quasi-static hydrodynamic limit in which the temperature
evolves isentropically and the cyclic pressure attractor is a narrow ellipse aligned with
the Navier--Stokes diagonal, whose width is set by $\omega\tau_\pi$. For fast
driving, $\omega\tau_\pi \gg 1$, the shear stress can no longer follow the rapidly
oscillating gradients and instead saturates at the free-streaming value. This
saturation, together with the associated suppression of viscous heating, is
correctly captured by the finite relaxation time of MIS but is missed entirely by
Navier--Stokes, which spuriously overestimates both the shear stress and the
heating, as seen in \cref{fig:MIS}. Throughout, the late-time cyclic attractor is accurately described by
linearised MIS theory, which asymptotes to the tilted ellipse of \cref{eq:ellipse_asymptote} whose width and tilt
grow with $\omega\tau_\pi$, as shown by the phase-space portraits in \cref{fig:MIS2}.

A central result, shown in \cref{fig:kinetic,fig:kinetic_ps}, is the universality of this attractor across the weakly coupled
kinetic theories: the RTA, scalar and Yang--Mills descriptions collapse onto one
another and onto the much simpler MIS reference, with only a small spread
developing at the largest frequencies. The minor deviations of the scalar theory
can be traced to its particle-number-conserving collision kernel, which allows a
chemical potential to develop during viscous heating. The strongly coupled
holographic response, by contrast, differs in two important respects. First, at low frequencies an
accurate description requires the second-order curvature coupling $\kappa$, which
is absent in the kinetic theories. As \cref{fig:holo,fig:holot_Tsub} show, MIS with $\kappa\neq 0$ tracks the holographic
result closely, whereas setting $\kappa=0$ leads to significant deviations that
are largest at the steep zero crossings. At larger frequencies, the holographic
response exceeds that of both MIS variants. Second, as is clearly visible in the phase-space portraits of \cref{fig:holo_ps}, the holographic
transient exhibits a pronounced ringing, which is associated with the non-vanishing real parts of the
frequencies of the
non-hydrodynamic quasinormal modes. This is in sharp contrast to kinetic theory, where for spatially homogeneous perturbations $(k=0)$ the singularities of the retarded response functions of the energy-momentum tensor that correspond to the non-hydrodynamic modes are located on the imaginary frequency axis~\cite{Romatschke:2015gic,Kurkela:2017xis}, corresponding to an exponential decay of transients.

Turning to large amplitudes in \cref{sec:nonlinear}, we find that non-linear effects become important. The enhanced viscous
heating induces a steady drift of the temperature, and the Navier--Stokes solution
overestimates this increase at large Knudsen number, while RTA, MIS and rBRSSS
continue to reproduce the temperature evolution well, see \cref{fig:nonlin_time}. The shear-stress evolution,
however, discriminates more sharply between the theories: the non-linear
second-order terms of rBRSSS are needed to follow the RTA trajectory, which it does
considerably better than MIS, as is evident from the phase-space trajectories in \cref{fig:nonlin_ps}. A systematic scan over amplitude and maximum
Knudsen number in \cref{fig:nonlin_scan} shows that the agreement between rBRSSS and RTA persists as long as
\emph{either} parameter remains small, and breaks down only once both become
large. The deformation of the attractor ellipse is controlled predominantly by
the amplitude rather than by the Knudsen number.

Finally, in \cref{sec:longtime} we have examined the long-time fate of the system under a sustained
drive. Because viscous heating continuously injects energy into the medium, the
average temperature grows over many cycles, the relaxation time $\tau_\pi$
decreases, and the effective drive strength $\omega\tau_\pi$ falls. This slow drift is fully captured
by the linearised MIS prediction once the dynamical temperature evolution is taken into
account, as demonstrated in \cref{fig:LargeTimesEvol}. As illustrated in \cref{fig:LargeTimes}, the decreasing $\omega\tau_\pi$ causes
the cyclic attractor to spiral inward from a wide, tilted ellipse onto the narrow
ellipse aligned with the Navier--Stokes diagonal, and the system to relax back
towards the Navier--Stokes regime at asymptotically late times.

Taken together, our results demonstrate that periodically driven 
systems provide a clean and versatile setting in which to study the emergence,
universality and eventual fate of hydrodynamic attractors far from the
Bjorken paradigm. The universality of the small-amplitude attractor across weakly
coupled theories, the distinguishing role of the curvature coupling and
quasinormal ringing at strong coupling, and the importance of non-linear
second-order terms at large amplitude all point to periodic driving as a sensitive probe
of the underlying microscopic dynamics. In the future, it would be interesting to extend the present study to
non-conformal equations of state~\cite{Spalinski:2025ngd,Romatschke:2017acs},
as well as to more realistic multi-dimensional expansion geometries~\cite{Romatschke:2017acs}.

\section*{Data availability}

The numerical data for the kinetic theory and holographic results, together with
the scripts that solve the hydrodynamic equations and produce all figures in this
paper, are publicly available at
\url{https://github.com/amazeliauskas/periodic_attractor_plots}.

\acknowledgments

The authors would like to thank J\"urgen Berges, Tilman Enss, Oscar Garcia-Montero, Lorenzo Gavassino, Michal Heller, Aleksi Kurkela, Jorge Noronha, Wilke van der Schee, Alexandre Serantes, Alexander Soloviev, Ismail Soudi and Clemens Werthmann for useful discussions.
A.M. and T.M. are supported by the Deutsche Forschungsgemeinschaft (DFG, German Research Foundation) through the Emmy
Noether Programme (project number 496831614) and CRC 1225 ISOQUANT (project number 273811115).
S.Schl., S.Schn., L.N.O. and M.V. are supported by the DFG
through the CRC-TR 211 `Strong-interaction matter under extreme conditions' -- project
number 315477589 -- TRR 211.

The authors used Claude (Anthropic) for assistance in developing and debugging simulation code, and for language editing of the manuscript. The authors verified all code and text and are solely responsible for the content.

\appendix
\crefalias{section}{appendix}
\section{Details of the holographic calculation}\label{app:holoappendix}

This appendix summarises the procedure used to solve the linearised bulk equation
\labelcref{Eq:gauge-invariant equation}, to determine the dynamical part of the retarded
correlator \labelcref{eq:GR_dyn} and the associated quasinormal frequencies and residues
entering \cref{Eq:Holo_GR}, and to construct the response functions
\labelcref{Eq:Gtt_holo,Eq:Longitudinal_StressCorrelator}.

\paragraph{Methodology:} We consider linearised metric perturbations
\labelcref{eq:holopertubation} at $k=0$ around the static black-brane background of
\cref{sec:holography_analytic}, and work in radial gauge, $h_{rM}=0$. Following
Ref.~\cite{Kovtun:2005ev}, the residual diffeomorphisms are removed by working with the
gauge-invariant combination $Z$, whose dynamics is governed by
\cref{Eq:gauge-invariant equation}. We solve this equation as a power series around
the horizon,
\begin{align}
    Z(r;\mathfrak w) = \sum_{i=0}^{N} c_i(\mathfrak w)\,(r-1)^i\,,
\end{align}
which implements regularity at $r=1$. Substituting this ansatz into
\cref{Eq:gauge-invariant equation} and demanding that the coefficient of
$(r-1)^{i-1}$ vanish determines every $c_{i\ge1}(\mathfrak w)$ recursively in terms of
$c_0,\dots,c_{i-1}$. We iterate this recursion up to $N=\mathcal O(10^2)$, which is
sufficient for the desired numerical accuracy on $r\in(0,1]$.

The horizon-regular series is then re-expanded around $r=0$,
\begin{align}
    Z(r;\mathfrak w) = Q(\mathfrak w)\big[1+\dots\big]
    + \Big[R(\mathfrak w) - Q(\mathfrak w)\,\mathfrak w^4 \log r\Big]\, r^4
    + \mathcal O(r^5)\,.
\end{align}
The logarithm reflects the conformal anomaly, in analogy with the logarithmic terms in
\cref{Eq:NearBdyExp,Eq:NearBdyExp1,Eq:NearBdyExp2}. Its coefficient is fixed by $Q(\mathfrak w)$ and carries no
independent information, so we subtract it before reading off the coefficient
$R(\mathfrak w)$ of the normalisable $r^4$ term. At the working truncation order,
$Q(\mathfrak w)$ and $R(\mathfrak w)$ are polynomials in $\mathfrak w$, whose ratio
enters \cref{eq:GR_dyn}.

\paragraph{Local contact term:}  
In addition to the dynamical contribution encoded in \(R/Q\), the renormalised stress tensor contains local terms constructed from the boundary metric. At \(k=0\), the frequency-dependent local contribution is quartic in \(\mathfrak w\), and the full correlator contains \(R/Q+(7/12)\mathfrak w^4\) in the minimal subtraction scheme of \cref{sec:Holography} (i.e., without finite counterterms). Because this contribution is analytic in \(\mathfrak w\), it does not modify the quasinormal frequencies, their residues, or the static limit. 

\paragraph{From the retarded correlator to the response function:}
The holographic retarded correlator is defined through the linear response of the tensor
density $\sqrt{-g}\,T^{\mu\nu}$ to the metric perturbation. In the convention of
Refs.~\cite{Policastro:2002tn,Kovtun:2005ev},
\begin{align} \label{eq:GR_response}
    \delta\langle \sqrt{-g}\,T^{\mu\nu}\rangle
    = -\frac{1}{2}\,G_R^{\mu\nu,\eta\eta}\,\delta g_{\eta\eta},
\end{align}
where the factor $-\tfrac12$ originates from the source coupling
$\tfrac12\int d^4x\, h_{\mu\nu}T^{\mu\nu}$ \cite{Policastro:2002tn}. The response functions in \cref{eq:linresponse}, are defined with respect to
the metric perturbation $\delta g_{\eta\eta}$, and for the tensor components $T^{tt}$ and
$T^\eta{}_\eta$ rather than for the tensor density. For $g_{\eta\eta}=a^2$ with
$a=1+\delta a$, the linear metric perturbation is $\delta g_{\eta\eta}=2\delta a$. In
addition, one must account for the difference between the variations of
$\sqrt{-g}\,T^{\mu\nu}$ and $T^{\mu\nu}$,
\begin{align}
    \delta\big(\sqrt{-g}\,T^{\mu\nu}\big) = \delta T^{\mu\nu}
    + \tfrac12\, T^{\mu\nu}_{(0)}\, \delta g_{\eta\eta}\,,
\end{align}
with $T^{tt}_{(0)} = e$ and $T^{\eta\eta}_{(0)} = P_\text{eq}$, as well as for the
conversion from $T^{\eta\eta}$ to $T^\eta{}_\eta$,
$\delta T^{\eta}{}_{\eta} = \delta T^{\eta\eta} + P_\text{eq}\, \delta g_{\eta\eta}$.
These contributions are proportional only to the equilibrium energy density and pressure
and therefore enter as frequency-independent contact terms. Combining these relations
with \cref{eq:GR_response},
\begin{align} \label{eq:GR_to_G}
    2G^{tt,\eta\eta} = -\left( G_R^{tt,\eta\eta} + e \right), \qquad
    2G^{\eta\eta,\eta\eta} = -\left( G_R^{\eta\eta,\eta\eta} - P_\text{eq} \right).
\end{align}
The contact terms of the retarded correlators are fixed at $k=0$ and $\omega\neq0$ by the
Ward identities derived in Ref.~\cite{Policastro:2002tn},
\begin{align} \label{eq:PSS_contact}
    G_R^{tt,\eta\eta} = P_\text{eq}\,, \qquad
    G_R^{\eta\eta,\eta\eta} = P_\text{eq} + c_s^2 (e + P_\text{eq})
    + G_R^{\eta\eta,\eta\eta}\Big|_{\rm dyn}\,,
\end{align}
where the static value $G_R^{\eta\eta,\eta\eta}(0) = \tfrac73 P_\text{eq}$ agrees with
the hydrodynamic limit of Ref.~\cite{Policastro:2002tn}. Inserting
\cref{eq:PSS_contact} into \cref{eq:GR_to_G} and including the local term above yields
\cref{Eq:Gtt_holo,Eq:Longitudinal_StressCorrelator}, whose energy response reproduces
the conservation equation and whose static stress response reproduces the adiabatic
pressure change $\delta P_\text{eq} = c_s^2\,\delta e$.

\paragraph{Quasinormal frequencies and residues:} The quasinormal frequencies
\labelcref{Eq:QNMcondition} are obtained as the zeros of the truncated polynomial
$Q(\mathfrak w)$. The lowest pair agrees with the known quasinormal spectrum of the
$\mathcal N=4$ black brane at $k=0$ \cite{Starinets:2002br}, providing a non-trivial check
of the series solution. Since $Q$ has a simple zero at each $\mathfrak{w}_n$, the dynamical part of the correlator has a simple pole there. To obtain the residue in the physical
frequency entering \cref{Eq:Holo_GR}, we regard $R$ and $Q$ as functions of
$\Omega$ via $\mathfrak{w}=\Omega/(2\pi T)$, such that
\begin{equation}
  \mathrm{res}_{\Omega_n}\!\left(\frac{R}{Q}\right)
  = \frac{R(\Omega_n)}{Q'(\Omega_n)}
  = 2\pi T\,\frac{R(\mathfrak{w}_n)}{\partial_{\mathfrak{w}} Q(\mathfrak{w}_n)}\,,
\end{equation}
where the prime denotes the derivative with respect to $\Omega$ and
$\Omega_n = 2\pi T\,\mathfrak{w}_n$. This gives the residue $\rho_n(\Omega_n)$
in \cref{Eq:Holo_GR} and, through \cref{eq:GeneralResponseLin}, the transient excitation amplitude
$\gamma_n(\omega_0)$ of each non-hydrodynamic mode shown in \cref{Fig:QNM spectrum}.

\bibliographystyle{JHEP}
\bibliography{citations}
\end{document}